\documentclass[a4paper,useAMS,usenatbib]{mn2e}

\usepackage[pdftex]{graphicx}
\usepackage{amssymb,amsmath}
\usepackage{booktabs}
\usepackage{afterpage}
\usepackage{pdflscape}
\usepackage{multirow}

\usepackage{array}

\usepackage{capt-of}

\usepackage{lipsum}

\title[\bf Galactic disc kinematics from A and F stars]{\bf A and F stars as probes of outer Galactic disc kinematics}
\author[A. Harris et. al.]{{\parbox{\textwidth}{A. Harris,$^{1}$\thanks{E-mail:a.harris7@herts.ac.uk} J. E. Drew$^1$, H. J. Farnhill$^{1}$, M. Mongui\'{o}$^{1}$, M. Gebran$^{2}$, N. J. Wright$^{3}$,\\ J. J. Drake$^{4}$, S. E. Sale$^{5}$}}
\\\\
$^{1}$School of Physics, Astronomy and Mathematics, University of Hertfordshire, College Lane, Hatfield AL10 9AB, UK\\
$^{2}$Department of Physics and Astronomy, Notre Dame University - Louaize, PO Box 72, Zouk Mikael, Lebanon \\
$^{3}$Astrophysics Group, Keele University, Keele ST5 5BG, UK\\
$^{4}$Harvard Smithsonian Center for Astrophysics, Cambridge, MA 02138, USA \\
$^{5}$Astrophysics Group, School of Physics, University of Exeter, Stocker Road, Exeter EX4 4QL, UK }
\begin{document}

\date{Accepted 2017 December 19. Received 2017 December 15; in original form 2017 October 27}

\pagerange{\pageref{firstpage}--\pageref{lastpage}} \pubyear{2017}

\maketitle

\label{firstpage}

\begin{abstract}
Previous studies of the rotation law in the outer Galactic disc have mainly used gas tracers or clump giants.  Here, we explore A and F stars as alternatives: these provide a much denser sampling in the outer disc than gas tracers and have experienced significantly less velocity scattering than older clump giants.  This first investigation confirms the suitability of A stars in this role.  Our work is based on spectroscopy of $\sim$ 1300 photometrically-selected stars in the red calcium-triplet region, chosen to mitigate against the effects of interstellar extinction.  The stars are located in two low Galactic latitude sightlines, at longitudes $\ell = 118^{\circ}$, sampling strong Galactic rotation shear, and $\ell = 178^{\circ}$, near the Anticentre.  With the use of Markov Chain Monte Carlo parameter fitting, stellar parameters and radial velocities are measured, and distances computed.   The obtained trend of radial velocity with distance is inconsistent with existing flat or slowly rising rotation laws from gas tracers \citep{BrandBlitz1993, Reid2014}.  Instead, our results fit in with those obtained by \citet{Huang2016} from disc clump giants that favoured rising circular speeds.  An alternative interpretation in terms of spiral arm perturbation is not straight forward.  We assess the role that undetected binaries in the sample and distance error may have in introducing bias, and show that the former is a minor factor.  The random errors in our trend of circular velocity are within $\pm 5$ km~s$^{-1}$.

\end{abstract}


\setlength{\extrarowheight}{5pt}
\section{Introduction}
\label{sec:introduction}



The kinematics of stars outside the Solar Circle is relatively poorly known. In addition to setting a constraint on the Galactic potential, secure knowledge of the motions of stars at all radii and longitudes in the Galactic disc facilitates studies of non-axisymmetric motion, such as streaming motions due to spiral density waves or transient winding arms. Understanding the kinematics of the disc in full, including the rotation law, will help us map out its structure, and set constraints on its formation and evolution. 

Within the disc of the Milky Way, the rotational motion is the dominant component. The IAU recommended standard of the circular velocity at the Sun is $220$\,km\,s$^{-1}$ \citep{KerrLyndenBell1986}. This result is based mainly on the analysis of tracers of the gaseous interstellar medium (ISM), including CO, H\,\textsc{i} 21\,cm and H\,\textsc{ii} radio recombination line observations. In recent decades the work of \citet{BrandBlitz1993} has been particularly influential. How the rotation varies outward from the Solar Circle is more challenging to measure than within because of the greater reliance on the uncertain distances to the ISM tracers. 

It has been common practice to associate CO and H\,\textsc{ii} radio recombination lines with star forming regions whose distances are specified by spectroscopic parallax. More recent studies have made use of masers in star forming regions with much better defined distances, and these have begun to favour a higher rotation speed. \citet{Honma2012} used a sample of 52 masers associated with pre-main sequence stars to find the rotational velocity of the local standard of rest (LSR) to be between $223$\,km\,s$^{-1}$ and $248$\,km\,s$^{-1}$, with a flat rotation curve. \citet{Reid2014} took a similar approach, using an expanded sample of over 100 masers to refine the circular rotation speed at the Sun to $240\pm 8$\,km\,s$^{-1}$. \citet{Bobylev2016} support a raised circular speed ($236\pm 6$\,km\,s$^{-1}$) and a nearly flat rotation curve, based on a large sample of open clusters. 

A feature of the studies to date is a drop in sensitivity with increasing Galactocentric radius, R$_{G}$, as the number or quality of available measurements declines. In the case of maser measurements, just a handful are available beyond R$_{G} \sim12$\,kpc. Similarly, the number of star clusters catalogued at R$_{G} \geq 10$\,kpc is limited to a few tens. This is an unavoidable consequence of the inherent low spatial frequency of these object classes outside the Solar Circle. Hence, Galactic rotation of the outer disc remains highly uncertain.

The frequency of stars at increasingly large Galactocentric radii becomes much higher than either masers or H\,\textsc{ii} regions and work has been undertaken exploiting helium-burning clump giants that can serve as a form of standard candle \citep{Castellani1992}.  Examples of this, emphasising measurement of the outer disc, are the works by \citet{Lopez-Corredoira2014}, using proper motions from PPMXL\footnote{PPMXL \citep{Roeser2010} is a catalog of positions and proper motions referred to the ICRS (International Celestial Reference
System)} and by \citet{Huang2016} using LSS-GAC\footnote{LAMOST Spectroscopic Survey of the Galactic Anticentre \citep{Xiang2017}. LAMOST is the Large Sky Area Multi-Object Fibre Spectroscopic Telescope.} and APOGEE \footnote{The Apache Point Observatory Galactic Evolution Experiment \citep[DR12 release, see][]{Alam2015}} radial velocities.    As older objects with ages exceeding 1\,Gyr, clump giants will be subject to significant kinematic scatter, including asymmetric drift.

In this study we examine another class of stellar tracer that can also provide a much denser sampling of the outer disc than is available from ISM tracers. We turn to A/F type stars. These stars offer the following advantages: they are intrinsically relatively luminous, with absolute magnitudes in the $i$ band of $\sim$0 to 3; as younger objects ($\sim100$ Myr), they have experienced significantly less scattering within the Galactic disc \citep{DehnenBinney1998}; and, as we shall show, the A stars especially are efficiently selected from photometric H$\alpha$ surveys. We are able to detect them at useful densities out to R$_{G} \sim$14\,kpc. Furthermore, the numbers of A type stars most likely exceed the numbers of clump giants in the outer disc, given that the outer disc is younger than the inner disc \citep{Bergemann2014}. Our own investigations of photometry from the INT  \footnote{Isaac Newton Telescope} Photometric H$\alpha$ Survey of the Northern Galactic Plane \citep[IPHAS,][]{Drew2005}, used here for target selection, supports this expectation.  

We explore radial velocity (RV) measurements of near-main sequence A/F stars as kinematic tracers in two outer-disc pencil beams, with a view to future fuller exploitation via spectroscopy on forthcoming massively multiplexed wide-field spectrographs (e.g WEAVE\footnote{WHT (William Herschel Telescope) Enhanced Area Velocity Explorer \citep{Dalton2016}}, in construction for the William Herschel Telescope). To greatly reduce the impact of extinction on the reach of our samples, our observations target the calcium triplet (CaT) region, in the far red part of the spectrum. Now is a good time to embark on this investigation in view of the imminent release of Gaia astrometry, which will include complementary proper motions of disc stars and support improved distance estimates.


We present a method of RV measurement in the CaT part of the spectrum and apply it to a dataset of $\sim1300$ spectra of A and F type field stars, in order to confirm their viability as tracers of Galactic disc kinematics and to pave the way for more comprehensive use of them in future. This first dataset is made up of stars in two pencil beams at Galactic longitudes $\ell=118^\circ$ and $\ell=178^\circ$. Our data were obtained at MMT using the multi-object spectrograph HectoSpec. Section \ref{sec:obsanddata} describes the observations obtained and how they were prepared for analysis.  In section \ref{sec:method} we detail the method used to obtain the RVs and stellar parameters and determine distances. The derived parameter distributions and error budgets are presented. The results obtained close to the anticentre at $\ell=178^\circ$, our control direction, demonstrate the validity of the analysis, while the results at $\ell=118^\circ$ indicate a strong departure from a flat or slowly-rising Galactic rotation law (section \ref{sec:results}). In section \ref{sec:discussion}, we review possible bias, make comparisons with earlier results and consider whether structure due to spiral arm perturbations may have been detected. Our results turn out to be consistent with the rising section of the rotation law presented by \citet{Huang2016} based on disc clump giants over the Galactocentric radius range 11 to 15\,kpc. A summary of our conclusions is given in section \ref{sec:conclusion}. 



\section{Observations and data}
\label{sec:obsanddata}
\subsection{Sightline and selection of targets for observation}
\label{sec:hec_data}

We chose to study two pencil-beam sightlines located at Galactic co-ordinates $\ell=178^{\circ}$, $b=1^{\circ}$ and $\ell=118^{\circ}$, $b=2^{\circ}$. These were chosen for the following reasons: 

\begin{itemize}
\item In principle, RVs measured at $l = 118^{\circ}$ should sample strong shear in Galactic rotation and hence provide stronger insight into how the rotation changes with Galactocentric radius outside the Solar Circle.  This particular sightline is also one that presents relatively low total extinction \citep{Sale2014} and limited CO emission \citep{Dame2001}, thereby promising access to greater distances.  It is consistent with these properties that there is a raised stellar density visible at this location of up to twice the average for the region \citep[see][]{Farnhill2016}.  This pencil beam most likely misses the main Perseus spiral arm, since the main belt of extinction and CO emission lies at higher latitudes than our chosen pointing by one to two degrees.  It is worth noting that the necessarily sparse H\,\textsc{ii} region data used by \citet{BrandBlitz1993} to measure the outer-disc rotation curve, revealed little change in apparent RV with distance in this location: all values cluster around an LSR value of $-$50 km s$^{-1}$. \citet{Russeil2003} reported a departure of $-21\pm10.3$\,km\,s$^{-1}$ from mean circular speed associated with the Perseus Arm.

\item Sightlines near the Galactic Anticentre intersect the direction of dominant circular motion essentially at right angles, leading to the expectation of measured RVs close to 0 (in the LSR frame) exhibiting negligible change with increasing distance.  We chose to observe an anticentre sightline so that it serves as a control directly revealing e.g. RV measurement bias and the magnitude of kinematic scatter.  Again, the particular choice made is of a pencil beam that is subject to relatively light extinction. At $l = 178^{\circ}$, higher extinction is found a degree away, on and below the Galactic equator.
\end{itemize}

The targets we observed have magnitudes spanning the range, $14.2 \leq i \leq 18.5$, which are expected to populate a heliocentric distance range of approximately 2 to 10\,kpc.


\begin{figure}
\centering
\includegraphics[width=\linewidth]{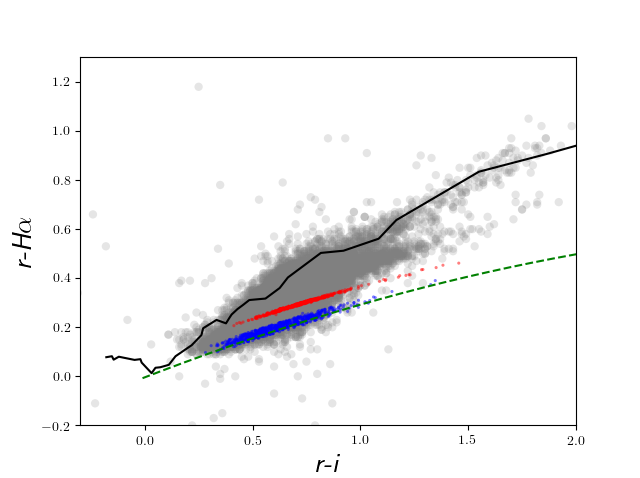}
\caption{\textit{IPHAS colour-colour diagram with HectoSpec targets overlaid. The grey points are IPHAS sources with $r<19$. The HectoSpec stars fall in two distinct selection strips: the top strip selecting F-type stars (red points) and the bottom strip selecting A-type stars (blue points). The black line is an empirical unreddened main-sequence track, and the green dashed line is the early-A reddening line. }}
\label{fig:colourselection}
\end{figure}

The sample was selected using the IPHAS $r-i$, $r-H\alpha$ colour-colour diagram, as shown in figure \ref{fig:colourselection}. \citet{Drew2008} described how this colour-colour diagram can be used to effectively select samples of near main sequence early-A stars. The candidate A stars were selected from a strip 0.04 mag wide, just above the early A reddening line shown in figure \ref{fig:colourselection}, and the candidate F stars were selected from a strip 0.08 - 0.09 mag above the same line. The use of both F and A stars ensure the heliocentric distance distribution is well-sampled, as the selected F stars will lie closer on average than the intrinsically brighter A stars. 


\subsection{Observations and selection of data for analysis}
The spectra were gathered using the MMT's multi-object spectrograph, HectoSpec \citep{Fabricant2005}, with the 600\,lines\,mm$^{-1}$ grating providing $\sim2.4\,\mbox{\AA}$ resolution as measured from narrow sky lines. The observations were performed over 6 nights in Sep-Nov 2011 and cover a wavelength range of $6532-9045\,{\mbox{\AA}}$ with a sampling of $0.56\,{\mbox{\AA}}$. Table \ref{tab:configs} details the observations on each date, giving the $i$ magnitude range targeted, exposure time, achieved count range, and number of targets. 

\begin{figure}
\centering
\includegraphics[width=\linewidth]{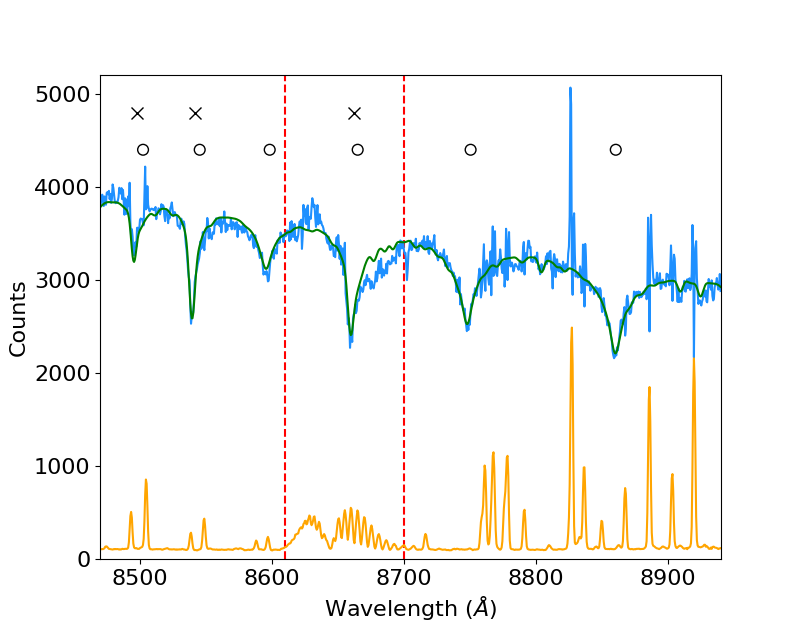}
\caption{\textit{Close up of the CaT region in the spectrum of an A star from the HectoSpec data set (blue) and the corresponding sky spectrum (orange). The latter has been scaled down to 15$\%$ of the original counts. The region between the red dashed lines is excluded from the MCMC full-parameter fitting procedure since this region is affected by sky lines which frequently do not subtract well. The green line is the MCMC template fit. The narrow emission-like features in the star spectrum are incompletely-subtracted sky lines.  The prominent $\sim$triangular absorption lines are Paschen lines, indicated by a circle. In later type stars, the CaT lines, indicated by a cross at 8498\,$\mbox{\AA}$, 8542\,$\mbox{\AA}$ and 8662\,$\mbox{\AA}$ become obvious, strengthening as the Paschen lines fade.}}
\label{fig:spectrum_closeup}
\end{figure}

The HectoSpec data were reduced using the HSRED\footnote{Information on the latest version of HSRED can be found at https://www.mmto.org/node/536} pipeline, and a mean sky spectrum is determined using dedicated sky fibres. This is subtracted off all target fibres in the same configuration. An example of a sky-subtracted spectrum of an A star (blue line) and its corresponding sky spectrum (orange line) can be seen in figure \ref{fig:spectrum_closeup}. The wavelength range shown ($8470-8940\,{\mbox{\AA}}$) is the region we use for measuring RVs and stellar parameters. Whilst much smaller than the total range covered by the data, it is chosen as it covers the CaT lines and some prominent Paschen lines. It is also relatively unaffected by telluric absorption lines. The only other strong photospheric feature potentially available to us is H$\alpha$, which is very far away in wavelength, and close to the blue limit of the MMT spectra.  The region between the red dashed lines in figure \ref{fig:spectrum_closeup} is excluded from further analysis since it is particularly dominated by bands of telluric emission which have proved difficult to subtract cleanly in many cases. The region also excludes the diffuse interstellar band at $8620\,{\mbox{\AA}}$.

\begin{table*}
\centering
\caption{The details of the observations: date of observation, $i$ magnitude range targeted, exposure time, achieved count range, number of targets observed, Galactic longitude of sightline.}
\label{tab:configs}
\begin{tabular}{cccccc}
\textbf{Date of observation} & \textbf{$i$ magnitude} & \textbf{Exposure (min)} & \textbf{Count range} & \textbf{Number of spectra} & \textbf{$\ell$ ($^\circ$)}     \\ \hline
17-09-2011                   & 16.5-17.5              & 140                     & 390-5163             & 266                        & \multirow{6}{*}{118$^{\circ}$} \\
18-09-2011                   & 16.5-17.5              & 135                     & 1000-4012            & 259                        &                                \\
19-09-2011 (a)               & $\leq$ 16.5            & 75                      & 204-2983             & 258                        &                                \\
19-09-2011 (b)               & 17.5-18.5              & 265                     & 143-2843             & 264                        &                                \\
21-10-2011                   & $\leq$ 17.5            & 135                     & 869-13003            & 256                        &                                \\
18-11-2011 (a)               & $\leq$ 16.5            & 75                      & 1452-12352           & 253                        &                                \\ \cline{6-6} 
17-11-2011                   & $\leq$ 18              & 165                     & 1192-4607            & 252                        & \multirow{2}{*}{178$^{\circ}$} \\
18-11-2011 (b)               & $\leq$ 17.5            & 120                     & 662-41194            & 262                        &                                \\ \hline
\end{tabular}
\end{table*}

\begin{figure}
\centering
\includegraphics[width=\linewidth]{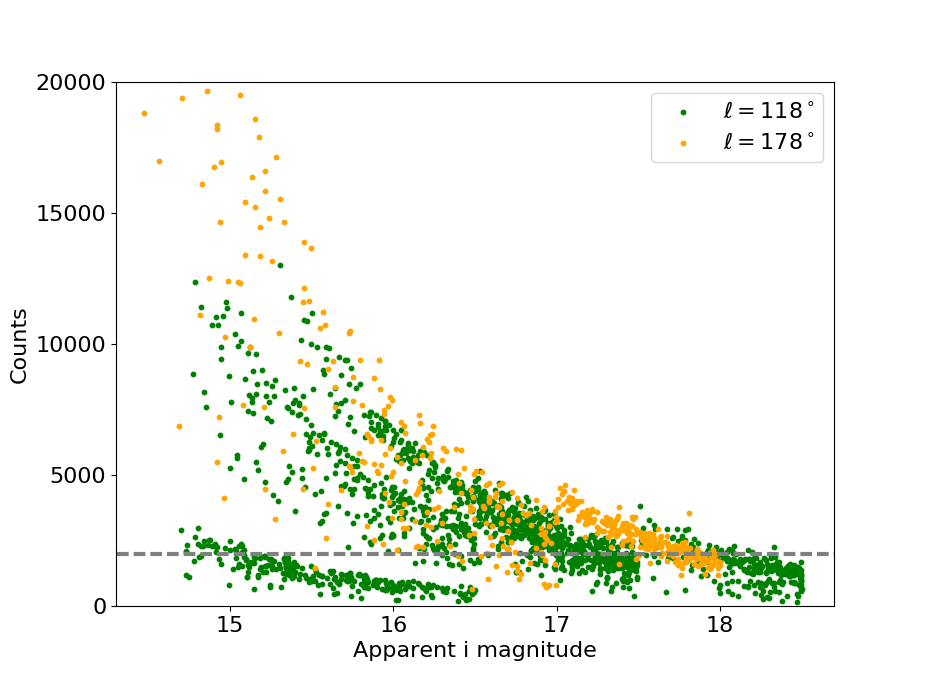}
\caption{\textit{Count level vs. apparent $i$ magnitude for $\ell=118^{\circ}$ (green) and $\ell=178^{\circ}$ (orange). The grey dashed line represents the minimum accepted count level of 2000. There are 749 stars with counts $<2000$, and 1321 with counts $\geq2000$.}}
\label{fig:count_vs_i}
\end{figure}

A quality cut was applied to the data, accepting spectra with an average count level of more than 2000 (between $8475-8675\,{\mbox{\AA}}$), to ensure a large enough signal-to-noise ratio (S/N) for reliable RV and stellar parameter measurements. The wavelength-averaged minimum S/N corresponding to this is 23. There are 887 spectra at $\ell=118^{\circ}$  and 434 spectra at $\ell=178^{\circ}$ that survive this cut. Figure \ref{fig:count_vs_i} shows the count levels vs apparent $i$ magnitude, with the horizontal line representing the minimum accepted count level. The distinct trends are due to weather, varying levels of moonlight, and exposure time changes (see table \ref{tab:configs}). The configurations exposed on 19-09-2011 and 18-11-2011, both for targets with $i \leq 16.5$, present contrasting count levels due to a significant transparency change. The effective magnitude faint limit is $i\sim17.5-18$.

HectoSpec spectra that show signs of red-leak contamination are betrayed by an upturn in the continuum at the red end reversing the decline seen shortward of $\sim 8600\,\mbox{\AA}$. These spectra are removed from the sample, leaving 855 target spectra in the $\ell=118^{\circ}$ sightline and 409 in the $\ell=178^{\circ}$ sightline. 

\section{Analysis of spectra}
\label{sec:method}
\subsection{Spectral fitting process}
\subsubsection{Overview}

RVs and the stellar parameters (effective temperature, T$_{eff}$, surface gravity, $\log{g}$, and rotational velocity, v$\sin{i}$) were measured using Markov Chain Monte Carlo (MCMC) full-parameter fits over the CaT range (8470-8940\,$\mbox{\AA}$). The target spectra are compared with a template set, interpolated as needed, by mapping the templates directly on to the observations and hence eliminating the need for separate continuum fitting to both the template spectrum and target spectrum.

The alternative method of RV measurement by cross-correlation was also explored. Each target spectrum was cross-correlated with every template, and the stellar parameters were adopted from the template which produced the tallest cross-correlation function peak. However, a significant weakness of this method is that it does not lend itself easily to error propagation. Consequently, the MCMC method has been favoured. The RV measurements for both methods are in agreement within the errors, with a slight bias to more negative values in the case of cross-correlation. The median difference between MCMC and cross-correlation RV measurements is 2.0\,km\,s$^{-1}$ -- an amount that falls inside the median RV error estimate of $\sim 4.4/6.8$\,km\,s$^{-1}$ for F stars/A stars.

\subsubsection{Template spectra}
\label{sec:templates}

In order to determine the needed quantities from the target spectra obtained, comparisons need to be made with a set of synthetic spectra. These synthetic spectra were calculated using the approach of \citet{Gebran2016} and \citet{Palacios2010}. These authors have used SYNSPEC48 \citep{HubenyLanz1992} to calculate the spectra based on ATLAS9 model atmospheres \citep{Kurucz1992}, which assume local thermodynamic equilibrium, plane parallel geometry and radiative and hydrostatic equilibrium. We collected 735 spectra that sample the parameter domain as follows:
\begin{itemize}
\item $\lbrack$Fe/H$ \rbrack = 0, -0.5$
\item $\lbrack \alpha$/Fe$\rbrack = 0$
\item $5000 \leq $ T$_{eff}$ (K) $ \leq 15000$, in steps of $500$\,K
\item $3.0 \leq \log{g} \leq 5.0$, in steps of $0.5$
\item $0 \leq $ v$\sin{i}$ (km\,s$^{-1}$) $\leq 300$\,km\,s$^{-1}$, in steps of 50\,km\,s$^{-1}$
\end{itemize}
The metallicity is not treated as a free parameter: instead we use two distinct template sets, one with $\lbrack$Fe/H$ \rbrack=0$ and the other $\lbrack$Fe/H$ \rbrack = -0.5$, and compare the results. We do this because of the limited spectral coverage and resolution of our data. Our numerical trials of metallicity as a free parameter showed it to be underconstrained and prone to interfere with the descent onto the values of other parameters. Nevertheless, there is an expectation that with increasing heliocentric distance along both pencil beams, there is likely to be a gradual decline in metallicity. \citet{Friel2002, Chen2003, Rolleston2000} found the metallicity gradient in the Milky Way to be $\sim -0.06$\,dex\,kpc$^{-1}$. At the median distances for the targets in each sightline, R$_G \sim 11$\,kpc for $\ell=118^{\circ}$ and R$_G \sim 13$\,kpc for $\ell=178^{\circ}$, this corresponds to $\lbrack$Fe/H$ \rbrack \sim -0.18$ and $\lbrack$Fe/H$ \rbrack \sim -0.29$ respectively. In contrast, \citet{Twarog1997, Yong2005} found $\lbrack$Fe/H$ \rbrack=-0.5$ at R$_G \geq$ 10\,kpc. Hence it is appropriate to gauge the effect of modest changes in adopted metallicity. 

The model spectra have a resolution of R $=10000$, cover $3600-9650\,{\mbox{\AA}}$ and have a constant wavelength sampling of $0.05\,\mbox{\AA}$. However for this work, the templates have been broadened with a gaussian filter  to match the resolution of the target spectra and rebinned to a sampling of $0.56\,{\mbox{\AA}}$ (also to match the data).

\subsubsection{MCMC assisted full-parameter fitting}
\label{sec:mcmc}

The parameters of astrophysical interest that we need to derive from each spectrum are T$_{eff}$, $\log{g}$, v$\sin{i}$ and RV. The first step in this process has to be a mapping of template spectra on to each observed spectrum. The mapping function we have adopted is linear which we find to be an adequate approximation over the short wavelength range considered. This introduces two extra free-parameters to be fit: the gradient and the intercept of the mapping function. This approach is preferred over the alternative of separate continuum fits to the template and observed spectra because it is then possible to monitor how the mapping numerically influences the outcome. 

The posterior probability distributions of the 6 free parameters (T$_{eff}$, $\log{g}$, v$\sin{i}$, RV, normalisation gradient and intercept) are obtained assuming a likelihood function of the form:
\begin{equation}
\mathcal{L}(\mu, \sigma^2 ; x_1, ..., x_n) \propto e^{-\frac{1}{\alpha}\sum\limits_{i}{\frac{(x_{i}-\mu_{i})^2}{2\sigma_{i}^2}}}
\end{equation}
where $x$ refers to the target star spectrum sampled at pixels $i=1...n$ and $\mu$ refers to the template spectrum. Since the pixels in a resolution element are not independent, we compensate by dividing the summation by the number of pixels per resolution element, $\alpha$. This is 4.3 in our case. 
$\sigma_i$ is the noise level of the target star at pixel $i$, given by
\begin{equation}
\sigma_i=\sqrt{\sigma_{sky+star, i}^2 + \sigma_{sky, i}^2}
\end{equation}
where $\sigma_{sky+star, i}^2$ is the count level of the raw spectrum (the sky and star together), and $\sigma_{sky, i}^2$ is the count level of the sky spectrum. This likelihood function assumes the target spectrum has gaussian errors.

By linearly interpolating the template grid, templates with intermediate parameter values are produced and hence the model parameter space is continuous. The RV range is sampled by shifting the wavelength axis of the template according to the Doppler formula. 

The `\textsc{emcee}' \textsc{python} package \citep{Foreman-Mackey2013} is used to execute the MCMC parameter space exploration. Our procedure is set up for 200 `walkers'. The priors used are flat for all parameters, with the range available to each parameter matching the range in  the template set. The range of the RV prior covers all realistic values ($\pm 500$\,km\,s$^{-1}$). The slope and intercept of the function mapping the template on to the target spectrum are nuisance parameters, for which the range on the prior was determined by experimentation with the dataset. After many steps the walkers converge (we use 2000 steps with a 700 step burn-in) and the distribution of parameter values returned by them define the posterior probability distribution of the parameters. These are typically of the form of a 2D gaussian. The medians of the marginalized distributions are adopted as the best estimates of parameter values and the uncertainties are based on the 16th and 84th percentiles. Figure \ref{fig:spectrum_closeup} shows an example of a template (green line) with best-estimate parameters for the observed spectrum (blue line). 

\subsection{Derived stellar parameters}

\subsubsection{T$_{eff}$}
\label{sec:teff}

\begin{figure}
\centering
\includegraphics[width=\linewidth]{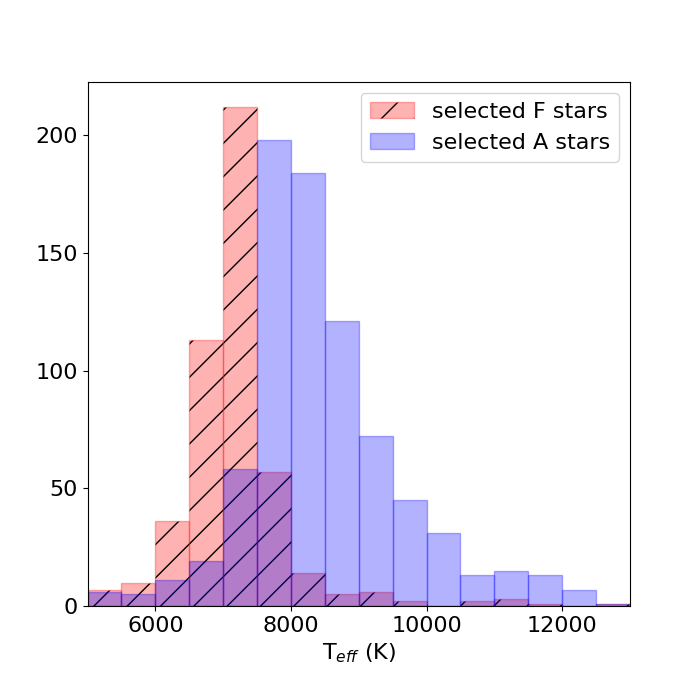}
\caption{\textit{Superposed distributions of measured T$_{eff}$ values (for both sight lines combined) of stars selected as probably F-type (red histogram) and A-type (blue histogram). Of the stars initially selected as A-type, only 12\% are cooler F/G stars.  }}
\label{fig:teff}
\end{figure}

The measured T$_{eff}$ distribution for both sightlines combined can be seen in figure \ref{fig:teff}. The red bars represent the stars originally selected from the IPHAS $r-i$, $r-H\alpha$ colour-colour diagram as F stars, and the blue bars represent the A stars. 78$\%$ of the stars selected as candidate F stars have measured T$_{eff}$ values of $6000-7500$\,K (typical of F stars), and $81\%$ of the candidate A stars have T$_{eff}$ values of $7500-10000$\,K (typical of A stars). It is necessary to point out that the template set spans temperatures inclusive of G stars (T$_{eff} \leq 6000$\,K) and B stars (T$_{eff} > 10000$\,K). Stars measured as G stars make up only $2\%$ of the total sample and B stars $4\%$. These are mainly stars that have contaminated the selected IPHAS $r-i$, $r-H\alpha$ regions. Given the modest levels of contamination (e.g only 12\% of the inital A star selection turned out to be cooler than 7500\,K), the method of selection is shown to be very effective and practically viable. For simplicity in the remainder of the paper, we label stars as either F or A-type, with T$_{eff} = 7500$\,K as the boundary dividing them. 

\begin{figure*}
\centering
\includegraphics[width=\linewidth]{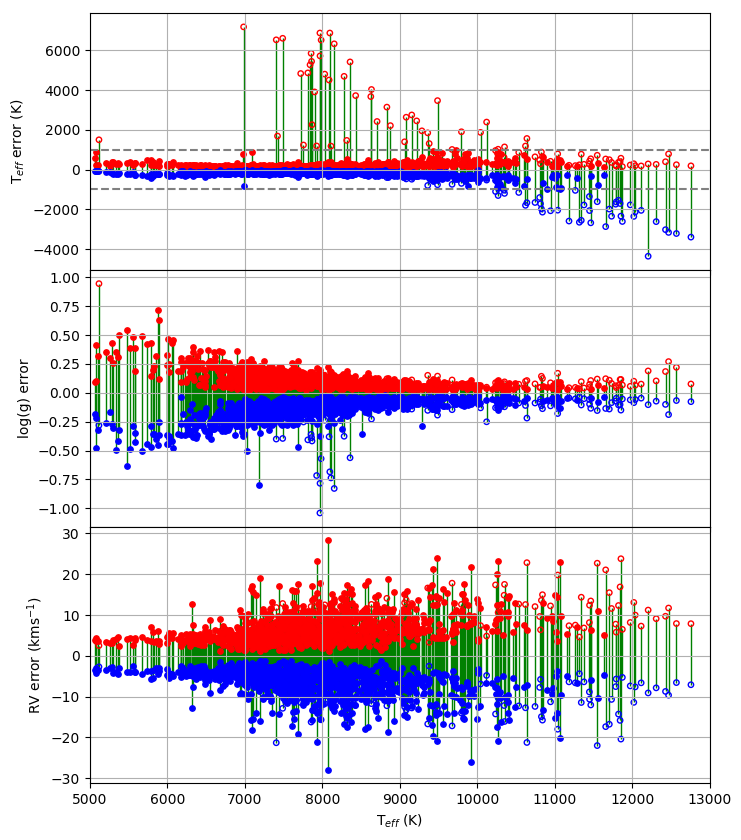}
\caption{\textit{Positive (red) and negative (blue) errors in T$_{eff}$ (top panel), $\log{g}$ (middle panel) and RV (bottom panel) as a function of T$_{eff}$. The grey dashed lines in the top panel represent the T$_{eff}$ error cut: points with $|error| >$1000\,K that are removed from the sample are shown in this figure, represented by empty circles. The green bars connect the positive and negative errors for each individual target. }}
\label{fig:errors}
\end{figure*}

The positive and negative errors in T$_{eff}$ as a function of T$_{eff}$ are shown in the top panel of figure \ref{fig:errors}. The median of the positive and absolute values of the negative errors combined is $\sim$ 150\,K however the average error increases with temperature and there are a number of targets that have large asymmetric errors, betraying an unresolved fit dilemma. As temperature increases, the Paschen line depths increase until T$_{eff} \sim$ 9000\,K, thereafter they become more shallow again. This means there can be a degeneracy where the Paschen line profiles of a lower temperature template are similar to that of a higher temperature template, with the addition of line broadening effects from the $\log{g}$ and v$\sin{i}$ parameter. This set of stars, with large T$_{eff}$ errors ($>1000$\,K), consist of only 85 of the 1261 targets, and are henceforth removed from the sample. These are represented by empty circles in figure \ref{fig:errors}.

Finally, this leaves a total sample made up of 705 A stars (T$_{eff} >$ 7500\,K) and 471 F stars (T$_{eff} \leq 7500$\,K): broken down into the sightlines, there are 473 A stars and 310 F stars in the $\ell=118^\circ$ sample, and 232 A stars and 161 F stars in the $\ell=178^\circ$ sample.  

\subsubsection{$\log{g}$}
\label{sec:logg}

The middle panel of figure \ref{fig:errors} shows the positive and negative errors in $\log{g}$ from the posterior distributions as a function of T$_{eff}$, for the A and F stars at solar metallicity. It is clear from the figure that the formal $\log{g}$ error rises steadily with decreasing effective temperature. The median error at T$_{eff} > 7500$\,K is 0.09, while below this it increases to 0.14.  This trend most likely tracks the growing importance of H$^-$ continuum opacity with decreasing effective temperature, causing the wings of the CaT lines to become less sensitive to surface gravity. \citet{Grey2009} have noted a `dead zone' among mid-late F stars in which dwarf and giant spectra are nearly indistinguishable.

In F stars with appreciably reduced Paschen line profiles, in particular, this underlying astrophysical trend is compounded to an extent by the moderate spectral resolution of the data. We find for these cooler stars that the fits begin to exhibit a 3-way degeneracy for combinations of temperature, gravity and metallicity.  An example of this degeneracy at work in the CaT in cooler F stars was presented by \citet{Smith1987}. However, with metallicity fixed, it is possible to identify temperature and gravity albeit with greater error on the latter. The returned F star $\log{g}$ distribution is skewed strongly in favour of near-MS objects, with a median value of 4.5 and interquartile range 4.1-4.8, tapering off into a tail reaching down to one object with $\log g \simeq 3.0$ (the lower bound on the template set).



\begin{figure}
\centering
\includegraphics[width=\linewidth]{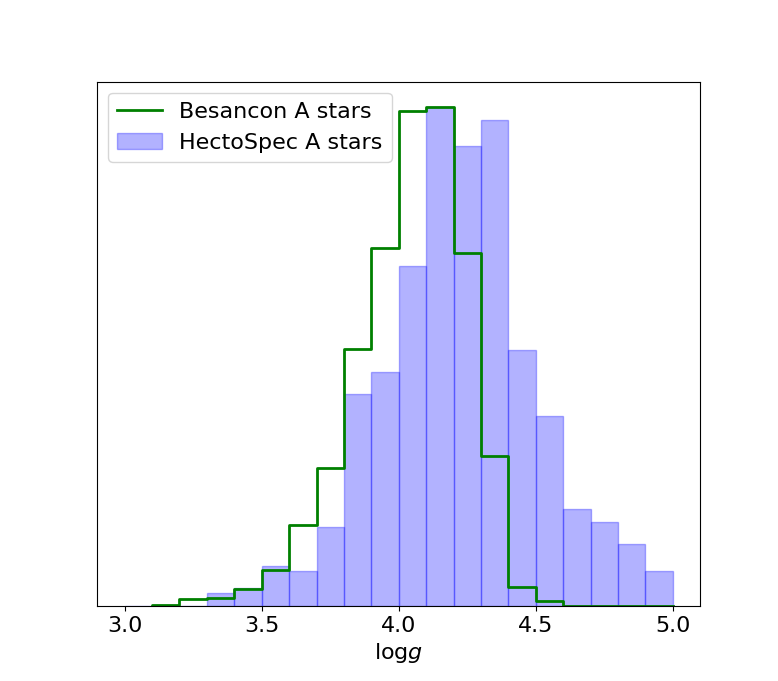}
\caption{\textit{Distribution of measured $\log{g}$ values of A stars for both sight lines combined (blue histogram). The distribution of $\log{g}$ values of A stars from a Besan\c con model for $\ell=118^\circ$ is also shown (green histogram). This has been convolved with a gaussian of $\sigma= 0.09$ to emulate the HectoSpec measurement error. }}
\label{fig:logg}
\end{figure}

The distribution of best-fit $\log{g}$ values for the A stars can be seen in figure \ref{fig:logg}. The peak of the distribution lies close to where we would expect it to be. Also shown is the distribution of $\log{g}$ for A stars from a Besan\c con model \citep{Robin2003} for $\ell=118^\circ$, after convolving with a gaussian of $\sigma= 0.09$ to emulate the HectoSpec measurement error, for a more useful comparison. The measured HectoSpec distribution has a tail at large values of $\log{g}$ that is not seen in the Besan\c con model, and its peak is not perfectly matched. A shift of the Besan\c con distribution of $+0.15$ brings the two distributions into rough alignment. This could be evidence of a bias in our measurements. We appraise the impact of this in section \ref{sec:comparison}. The stars occupying the high-end tail, which appears more extended than in the Besan\c con model distribution, are mainly cooler objects carrying larger-than-median errors (see figure~\ref{fig:errors}).


\subsubsection{v$\sin{i}$}

\begin{figure}
\centering
\includegraphics[width=\linewidth]{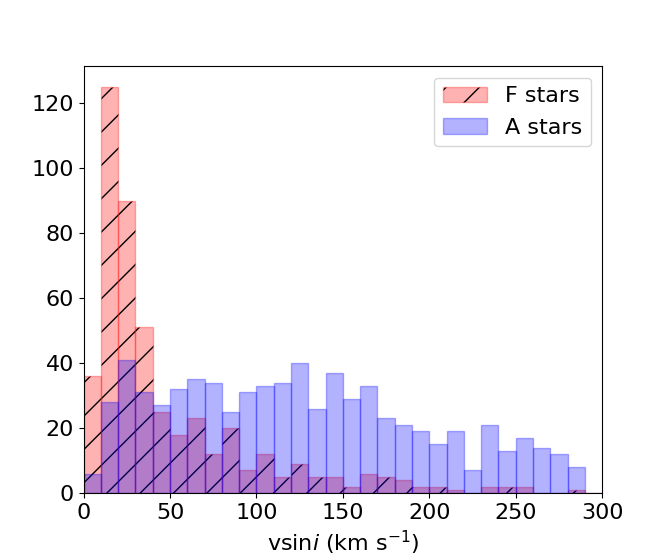}
\caption{\textit{The distributions of measured v~$\sin{i}$ values, separated into F stars (red) and A stars (blue), and shown overplotted. The two lines of sight are merged within each distribution. }}
\label{fig:vsini}
\end{figure}

Figure \ref{fig:vsini} shows the measured v$\sin{i}$ distribution for both sightlines combined, separated into F stars (red bars) and A stars (blue bars). The distribution is as expected, with generally low values for F stars and a spread from low to high values for A stars \citep{Royer2014}. The median error on v$\sin{i}$ is $\sim 20$\,km\,s$^{-1}$, increasing to $\sim$40\,km\,s$^{-1}$ for T$_{eff} > 10000$\,K - as expected for stars that are more commonly fast rotators. Unsurprisingly, there is evidence of a slight negative correlation between individual $\log{g}$ and v$\sin{i}$ parameter fits.

\subsubsection{RV}
Figure \ref{fig:errors} shows the errors on RV as a function of T$_{eff}$. The median error is $\sim 4.4\,$km$\,$s$^{-1}$ for F stars and $\sim 6.8$\,km\,s$^{-1}$ for A stars. This difference is attributable to the growing contrast of the CaT lines with decreasing effective temperature. The Solar motion needed to convert from the heliocentric to the LSR frame is taken as $(U,V,W) = (-11,+12,+7)$\,km\,s$^{-1}$ \citep{Schonrich2010}.

\subsection{A comparison with higher-resolution long-slit spectra}
\label{sec:isis}
As a check on the reliability of the derived parameters and their errors, we obtained additional long-slit red and blue spectra for 7 HectoSpec targets -- the positions and apparent $i$ magnitudes of which are shown in table \ref{tab:isis}. These were accompanied by 5 RV standard stars (with 3 observed more than once).  Of particular interest in this comparison are the measured surface gravities and RVs.  The 7 objects reobserved are spread in $\log{g}$ from 3.5 to 4.5, as determined from the HectoSpec data, lying across the peak of the distribution in figure \ref{fig:logg}. The T$_{eff}$ range, again determined from the HectoSpec data, spans $\sim 7900-9500$\,K, with one object outside this range with $\sim 11900\pm_{2600}^{150}$\,K -- the large error indicating the unresolved fit dilemma described in section \ref{sec:teff}.

\begin{table}
\centering
\caption{Positions and apparent $i$ magnitudes of the objects used for the HectoSpec-ISIS comparison.}
\label{tab:isis}
\begin{tabular}{ccccc}
\textbf{Target} & \textbf{RA} & \textbf{DEC} & \textbf{$i$ mag} & \textbf{ISIS red T$_{eff}$ (K)} \\ \hline
1               & 00:03:41    & 64:29:43     & 14.94                     & 7665$\pm_{101}^{115}$           \\
2               & 00:06:36    & 64:22:38     & 14.89                     & 7703$\pm_{73}^{76}$             \\
3               & 00:03:47    & 64:17:12     & 14.77                     & 7769$\pm_{95}^{89}$             \\
4               & 00:07:37    & 64:07:51     & 14.78                     & 8331$\pm_{219}^{2957}$          \\
5               & 00:05:36    & 63:57:55     & 14.84                     & 11043$\pm_{1332}^{344}$         \\
6               & 00:04:02    & 64:29:49     & 14.91                     & 11299$\pm_{1710}^{158}$         \\
7               & 00:07:13    & 64:46:18     & 14.82                     & 11370$\pm_{378}^{202}$       \\
\hline
\end{tabular}
\end{table}

These spectra were gathered as service observations using the Intermediate-dispersion Spectrograph and Imaging System (ISIS) of the 4.2\,m William Herschel Telescope during 3 nights in Oct-Dec 2016. ISIS is a high-efficiency, double-armed, medium-resolution (8-121\,$\mbox{\AA}$/mm) spectrograph. The R1200B grating was used on the blue arm, and the R1200R grating on the red arm, providing resolution elements for a 1" slit of 0.85\,$\mbox{\AA}$ and 0.75\,$\mbox{\AA}$ respectively. The wavelength coverage of the blue spectra is 3800-4740\,$\mbox{\AA}$ and of the red spectra is 8110-9120\,$\mbox{\AA}$. Both the blue and red spectra have a constant wavelength sampling of 0.22\,$\mbox{\AA}$ and 0.24\,$\mbox{\AA}$ respectively.

The raw images were processed and sky-subtracted and the resultant spectra was wavelength calibrated, all with use of the Image Reduction and Analysis Facility (\textsc{iraf}). The spectra were then passed through the MCMC full-parameter fitting method. For the red spectra the same wavelength limits were adopted as for HectoSpec, and for the blue the wavelength range used was $4000-4600\mbox{\AA}$, covering some of the Balmer series whilst excluding the Ca\,\textsc{ii} K and H lines since they suffer from interstellar absorption. The measured parameters and RV were then compared to those measured from the HectoSpec spectra, or to values from the literature in the case of the RV standards, to reveal any systematic differences. A method check was performed by comparing the measured parameters from the red and blue ISIS spectra, eliminating the potential for differences due to a change of instrument. The weighted mean difference between measured RV for the ISIS RV standard stars and RV from the literature is $-0.7$\,km\,s$^{-1}$ for the blue spectra and $+0.1$\,km\,s$^{-1}$ for the red spectra. Hence we are confident the ISIS wavelength scale is reliable. 

\begin{figure*}
\centering
\includegraphics{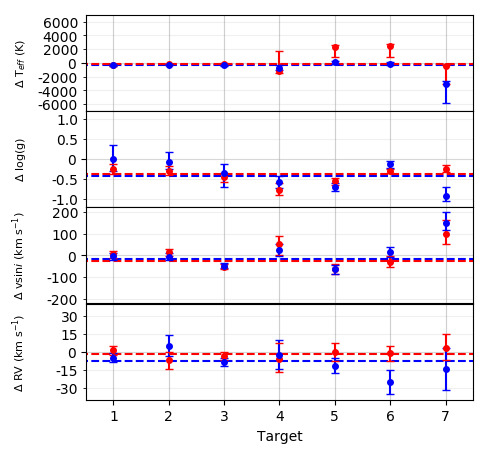}
\caption{\textit{Differences between ISIS and HectoSpec measured stellar parameters. The blue points represent the differences between measurements from the blue ISIS and HectoSpec spectra, and the red points specify differences between the red ISIS and HectoSpec spectra. The targets are in ascending order of T$_{eff}$, as determined from the ISIS red spectra. The error shown on each datapoint is the quadratic sum of the HectoSpec error and ISIS error. The dashed lines represent the weighted mean difference between the ISIS red and HectoSpec measurements (red line) and ISIS blue and HectoSpec measurements (blue line). }}
\label{fig:isis-comparison}
\end{figure*}

Figure \ref{fig:isis-comparison} shows the difference between the ISIS measurements (blue points for blue spectra and red points for red spectra) and the HectoSpec measurements. The targets are in ascending order of T$_{eff}$, as determined from the ISIS red spectra. In general the differences in outcome are reassuringly modest and show the method is working satisfactorily. The dashed lines represent the weighted mean difference between the ISIS red and HectoSpec measurements (red line) and ISIS blue and HectoSpec measurements (blue line). The weighted mean difference between the red ISIS and HectoSpec measurements, and the standard deviation of the spread, are: $\Delta$T$_{eff} = -215\pm1335$\,K, $\Delta \log{g} = -0.36\pm0.18$, $\Delta$v$\sin{i} = -24\pm53$\,km\,s$^{-1}$, $\Delta$RV $= -1.9\pm3.5$\,km\,s$^{-1}$. The small mean difference in measured RV indicates the HectoSpec wavelength calibration is systematically offset by an amount well below measured random errors. The larger difference in $\log{g}$ suggests there may be a bias towards larger values in the HectoSpec data. This potential bias is not the first evidence: the comparison between our A star $\log{g}$ distribution and that from a Besan\c con model is shown in section \ref{sec:logg} and we see the HectoSpec distribution peaks at a slightly larger value. We return to the impact of this possible bias in section \ref{sec:comparison}.

The blue data points in figure \ref{fig:isis-comparison} provide the comparison between the ISIS blue and HectoSpec data. The weighted mean offsets and standard deviations obtained are: $\Delta$T$_{eff} = -289\pm1017$\,K, $\Delta \log{g} = -0.43\pm0.32$, $\Delta$v$\sin{i} = -18\pm63$\,km\,s$^{-1}$, $\Delta$RV $= -7.7\pm8.8$\,km\,s$^{-1}$. As with the red, there is a sizable offset in $\log{g}$, suggesting the HectoSpec data may carry a positive bias. The RV difference is larger than that measured for the red spectra, potentially as a result of the fewer absorption lines fitted in the blue region and hence lower accuracy.

We have also reviewed the differences between parameters derived from the ISIS blue measurements and the ISIS red and HectoSpec data. \citet{Hales2009} performed a similar comparison between the red and blue ranges as observed with ISIS, and found a tendency for the red determined spectral type to be earlier by 0.9 subtypes. They attributed this to strengthening of the Ca\,\textsc{ii} H and K absorption lines by an interstellar component. Although we exclude these lines in the fitting procedure, we find a modest offset in effective temperature, with the red suggesting a slightly hotter star of 112\,K with very little change in surface gravity (0.02 dex).

The comparison with higher-resolution spectra has provided useful insights -- the offsets between the ISIS and Hectospec measurements are reassuringly modest and are similar to those expected due to random errors. Our employed method is working well. The larger offset in the derived surface gravities suggest a bias could be present to which we will continue to pay attention.

\subsection{Extinction, absolute magnitudes and distance modulii}
\label{sec:mag_ex_dis}

The distance modulus, $\mu$, of each star is calculated via the following equation
\begin{equation}
\mu = m_i - A_i - M_i
\end{equation}
where m$_i$ is the apparent magnitude in the $i$ band, A$_i$ is the extinction and $M_i$ is the absolute magnitude. The absolute magnitudes used are from Padova isochrones, interpolated with T$_{eff}$ and $\log{g}$ scales \citep{Bressan2012, Chen2015}. We choose a value for $M_i$ on the basis of the median $\log{g}$ and T$_{eff}$ returned by the fits. Where $\log{g}$ exceeds the maximum present in the Padova isochrones for the specified T$_{eff}$, we reset $\log{g}$ to this value. The median error on the absolute magnitudes (due to stellar parameter uncertainties) is $\sim$0.3. This is the dominant error source in our final results.

The extinction of each target was calculated using
\begin{equation}
A_i = 2.5 \lbrack (r-i)_{obs}-(r-i)_{int} \rbrack
\end{equation}
where $(r-i)_{obs}$ is the observed colour of the star, taken from IPHAS photometry, and $(r-i)_{int}$ is the intrinsic colour of the star. The intrinsic colours are calculated from the template grid via synthetic photometry and the value for each target star is interpolated on this grid based on the measured T$_{eff}$ and $\log{g}$. The coefficient of 2.5 is the ratio of A$_i$ to A$_r-$A$_i$ for main-sequence A/F stars with reddening levels similar to the HectoSpec data. We use the Fitzpatrick law with R$_V=3.1$. The median error on intrinsic colour (due to stellar parameter uncertainties) is $\sim$0.01, and on A$_i$ (due to stellar parameter and photometric uncertainties) is $\sim$0.05.  After dereddening the observed magnitudes, $\mu$ is obtained: across the entire sample the interquartile range for the error in $\mu$ is 0.2 to 0.4. The smallest errors are associated with early A stars (median error $\sim 0.2$).

\begin{figure}
\centering
\includegraphics[width=\linewidth]{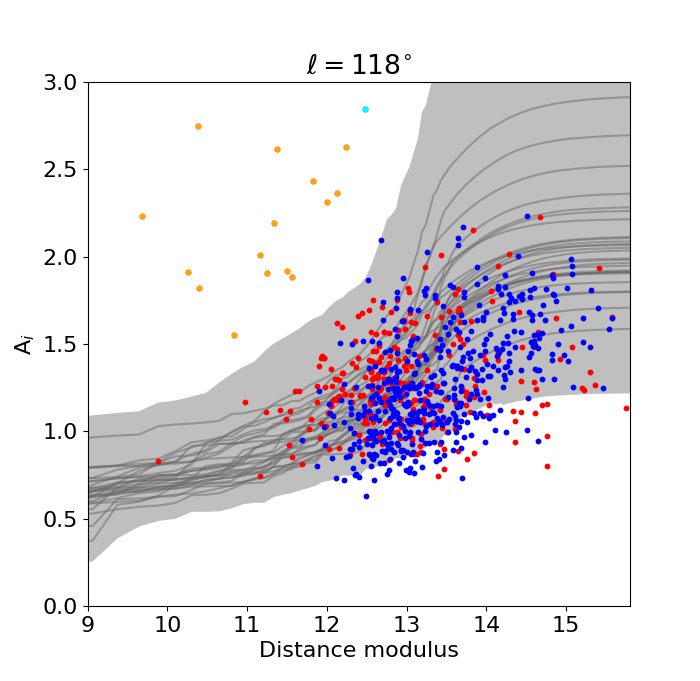}
\includegraphics[width=\linewidth]{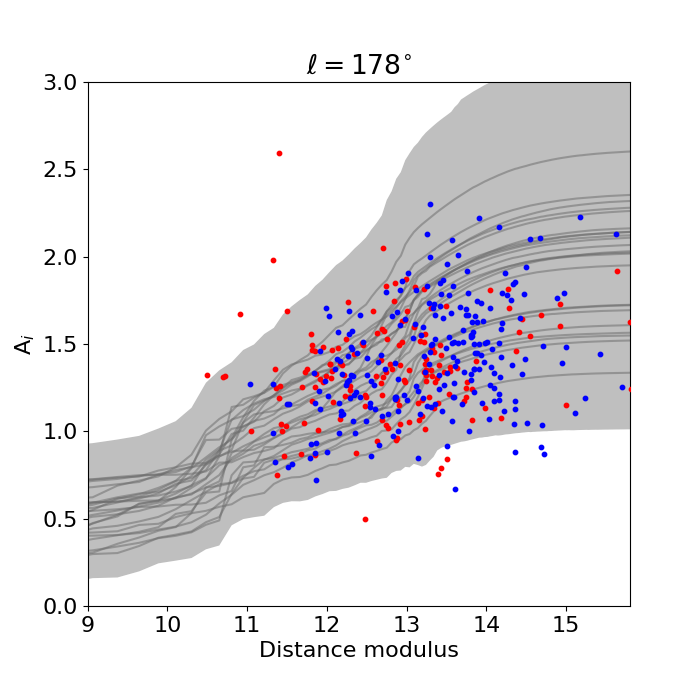}
\caption{\textit{The extinction, A$_i$, of F stars (red points) and A stars (blue points) as a function of distance modulus. Top: $\ell = 118^{\circ}$, bottom: $\ell = 178^{\circ}$. Also shown in each panel are the photometrically-predicted mean extinction trends (grey lines) across the pencil beam \citep{Sale2014}. The grey shaded region defines the expected dispersion in extinction at every distance within the beam. Stars that lie far from the trend in the upper panel are highlighted by a change of colour -- orange for the F stars and cyan for one A star. This colouring is preserved in figure \ref{fig:rv_vs_dist_118}.  }}
\label{fig:extinction}
\end{figure}

Figure \ref{fig:extinction} shows the A$_i$ extinctions as a function of estimated distance modulus for both sightlines. Also shown are the mean extinction trends due to \citet{Sale2014} across the two pencil beams, including the expected dispersion in extinction (grey-shaded region). This comparison provides some insight into the plausibility of the distance modulus distribution of our sample.  An important difference to be aware of is that the \citet{Sale2014} trends were computed from IPHAS photometry of all probable A-K stars, down to apparent magnitudes that are appreciably fainter (to $i \sim 20$) than those typical of our analysed spectra ($i < 18$). In both sight lines, there is evidence that the spectroscopic samples favour lower extinctions than the fainter-weighted photometrically-based trends. This is most likely a straight-forward selection effect. 

To check this, we have examined the impact of the counts cut we placed on the spectra included in the analysis: specifically, we have compared median estimates of extinction, derived from their $r-i$ colour, for the stars analysed (counts $> 2000$), with those for lower-count stars not analysed.  We find that for both directions the cut biases the typical extinction of the A stars (inferred from the available $r-i$ colours) to lower values: the strength of effect is that the unanalysed objects have a median A$_i$ that is greater by $0.4$ magnitudes. This helps explain the tendency for the blue A-star data points, particularly, to sit lower in figure~\ref{fig:extinction}.  Among the cooler, on-the-whole closer, F stars there is little difference.   The alternative explanation for this -- overlarge distance moduli -- runs into difficulty when it is recalled that there may be systematic overestimation of surface gravities (see sections \ref{sec:logg} and \ref{sec:isis}). We view this comparison as tensioning against accepting and correcting for such a bias in $\log{g}$, as this would drive up the distance moduli, creating a yet bigger offset.

\subsection{The effect of metallicity}
\label{sec:met1}

So far all the parameters obtained and described have been computed for solar metallicity. 

Stellar parameters and RVs have been measured again, as described in section \ref{sec:mcmc}, but with $\lbrack$Fe/H$ \rbrack$ set to -0.5 for the A stars. A new Padova absolute magnitude scale, suitable for the changed metallicity, was used for determining distance moduli. The reduced-metallicity fits typically return cooler temperatures ($\Delta $T$_{eff} \sim-400$\,K). The $\log{g}$ values are also lower ($\Delta \log{g} \sim-0.17$), partly compensating for the cooler temperatures in the estimation of distance.  The net effect on the distance modulus scale is a small increase compared with the solar metallicity scale ($\Delta \mu \sim+0.12$). The RVs measured adopting $\lbrack$Fe/H$ \rbrack = -0.5$ are slightly more negative: the median difference is $\sim-3.0$\,km\,s$^{-1}$.  

In the case of the F stars, there is a growing degeneracy between T$_{eff}$, $\log{g}$ and metallicity (see section \ref{sec:logg}). In a trial of $\lbrack$Fe/H$\rbrack=-0.5$ in fitting the F stars, we found the returned gravities to be unrealistically low for a population of objects that is more localised than the A stars whilst sharing the same faint magnitude limit. Moreover, it is highly improbable that many among the F star sample would present $\lbrack$Fe/H$\rbrack$ significantly less than 0, since the great majority of these fainter objects should lie within a distance of 5 kpc.    We estimate this limiting radius for the case of a warmer main sequence F star with T$_{eff} = 7000$\,K, $i = 17.5$ (see figure~\ref{fig:count_vs_i}), $A_i \sim 1.3$ (typical HectoSpec value -- see figure~\ref{fig:extinction}) and $M_i = 2.6$.  
At $\ell = 118^{\circ}$, a distance of 5\,kpc corresponds to a Galactocentric radius of $\sim$10.6~kpc, or a metallicity change of $\sim0.2$. Consequently throughout this paper we will fix the metallicity of the HectoSpec F stars at $\lbrack$Fe/H$\rbrack=0$. 



\section{The sightline radial velocity trends}
\label{sec:results}


The trend of RV with distance modulus can be compared with what we would expect to see as an averaged trend if consensus views of the Galactic rotation law apply. If all the stars move on circular orbits about the Galactic centre, we would expect to observe a trend of RVs that are a spread (due to velocity dispersion) around a curve whose shape depends on the sightline observed, the assumed rotation curve and LSR parameters (circular velocity and Galactocentric distance of the LSR, V$_0$ and R$_0$). The comparisons made in this section are with a flat rotation curve with LSR parameters R$_0=8.3$\,kpc, V$_0=240$\,km\,s$^{-1}$, and with the slowly rising rotation curve derived in \citet{BrandBlitz1993} with their adopted parameters  R$_0=8.5$\,kpc, V$_0=220$\,km\,s$^{-1}$. We begin with the results for $\ell=178^\circ$, the control direction, where the rotational component of motion of the stars is effectively nullified.



\subsection{$\ell=178^{\circ}$}
\label{sec:180}
\begin{figure*}
\centering
\includegraphics[width=\textwidth]{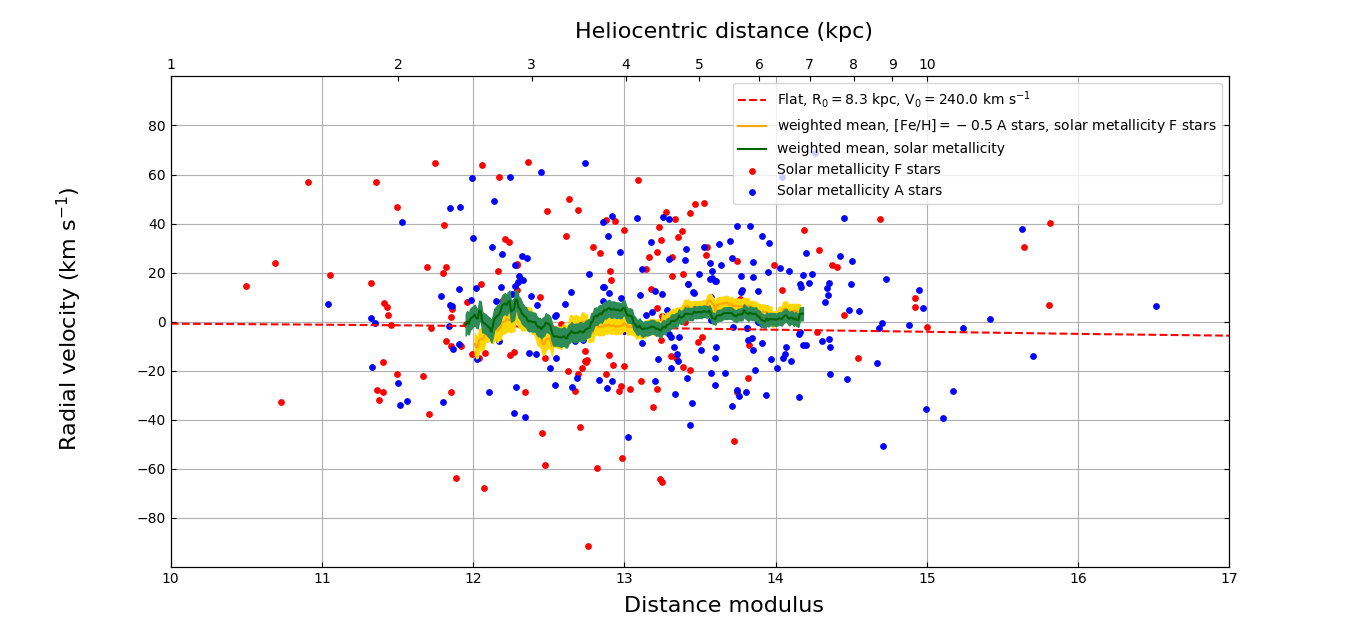}
\caption{\textit{The trend of RVs with distance modulus for $\ell=178^{\circ}$. The green line is the weighted mean of the RVs of the solar metallicity stars, and the yellow line is obtained on combining $\lbrack$Fe/H$ \rbrack = -0.5$ A stars with solar metallicity F stars. The shaded regions represents the standard error of the mean in RV. The blue points are A stars and the red points F stars, both with metallicity assumed as solar. The red dashed line shows the expected trend for a flat rotation law, which at this sightline will fall close to $\sim$zero regardless of the V$_0$, R$_0$ or shape of the rotation law adopted.}}
\label{fig:rv_vs_dm}
\end{figure*}

The trend of RV with $\mu$ for this pencil beam is shown in figure \ref{fig:rv_vs_dm}. The red dashed line shows the expected average trend based on a flat rotation curve (although at this sightline the shape of the rotation curve makes little difference). The green line is the weighted mean of the RVs for the solar metallicity stars, plotted as individual data points, at each step in $\mu$. The yellow line is the weighted mean for A stars with $\lbrack$Fe/H$\rbrack=-0.5$ and F stars with solar metallicity. The weighted RV is calculated via the following process:

\begin{itemize}
\item Every ($\mu$, RV) data point is assigned a normalised gaussian error distribution with $\sigma=\sigma_\mu$ where $\sigma_\mu$ is the error on $\mu$. 
\item At the i$^{th}$ step ($\mu=\mu_i$), the data points ($\mu_n$, RV$_n$) with error distributions in $\mu$ overlapping $\mu_i$ are included in forming the running average, provided $|\mu_n-\mu_i| < \sigma_{\mu_n}$. The weighting per data point is in proportion to the value of its gaussian error function at $\mu_i$ -- thereby taking into account the error in $\mu$ in all data points.
\item In obtaining the mean RV at $\mu=\mu_i$, the contributing RV values are also weighted in proportion to their errors. 
\item A final weighting is applied in computing the mean in order that the contribution from data points at $\mu < \mu_i$ balances the contribution from data points at $\mu > \mu_i$. This limits the influence of data from the most densely populated part of the $\mu$ distribution on the mean trend.
\item The minimum number of points required for calculation of the mean to either side of $\mu = \mu_i$ is 50. 
\end{itemize}
 

The shaded region around the mean trend line represents its error -- the standard error of mean RV at each $\mu_i$.

In figure \ref{fig:rv_vs_dm}, the F stars dominate the lower end of the range in $\mu$, while the A stars mostly occupy a spread from $\mu \sim 12$ to $\mu \sim 15$ (or a heliocentric distance range from 2 to 10\,kpc). Of the F stars, 90$\%$ (145 of 165) lie within $\mu= 11.4 - 14.5$ (d$=1.9-7.8$\,kpc). 90$\%$ of the A stars with $\lbrack$Fe/H$ \rbrack = 0$ (209 of 232) lie within $\mu=11.8 - 14.8$ (d$=2.3-9.0$\,kpc), and 90$\%$ of the A stars with $\lbrack$Fe/H$ \rbrack = -0.5$  (130 of 144) lie within $\mu=11.6 - 15.1$ (d$=2.1-10.6$\,kpc). \citet{Ruphy1996} found a radial cutoff of the stellar disc at R$_{G} = 15\pm2$\,kpc, and \citet{Sale2010} found the stellar density of young stars declines exponentially out to a truncation radius of R$_{G}=13\pm0.5_{statistical}\pm0.6_{systematic}$\,kpc, after which the stellar density declines more sharply. The density of our sample in this sightline also drops off at these distances, but at this stage we cannot be sure whether this decline originates in the Galactic disc or is a selection effect. 

We find the overall trend in RV follows the expected flat behaviour, although it is offset in velocity by a mean amount of $3.5$\, km\,s$^{-1}$. The results adopting $\lbrack$Fe/H$ \rbrack =-0.5$ for the A stars are very similar as expected, but with a slightly smaller mean offset of $3.0$\, km\,s$^{-1}$. There was evidence of a possible small wavelength calibration offset arising from the comparison of the HectoSpec observations with independent higher-resolution spectra, effecting the RV scale by only a couple of km\,s$^{-1}$, as described in section \ref{sec:isis}. Taking these findings together we infer that the RV scale is reliable to within $\sim 5$\, km\,s$^{-1}$ but cannot rule out the presence of a small positive bias in the measurements.



We have computed the RV dispersion of the sample stars, in broad spectral type groups, around the measured mean trend, in order to compare them with the expected dispersions from \citet{DehnenBinney1998} (see table \ref{tab:veldisp}). Clearly the measured dispersions are larger than the \citet{DehnenBinney1998} results. Part of this discrepancy is attributable to our measurement errors, but the larger share of it may be a consequence of stellar multiplicity. At least a half of the sample objects are likely to be members of multiple systems, and of those around 15\% may be in nearly equal mass ratio binaries \citep{Duchene2013}. Whilst the spectroscopic resolution of our data is insufficient to pick out spectroscopic binaries, there can nevertheless be binary orbital motions present up to a level of $\sim 40$\, km\,s$^{-1}$ (depending on phase) in typical cases (for total system masses of $3M_{\odot}$, with period P $\sim40$\, days). As the data in table \ref{tab:veldisp} indicate, the excess dispersion on top of measurement error is in the region of $10$\, km\,s$^{-1}$ for all 3 spectral type groups. We look at this in more detail in section \ref{sec:discussion}.

\begin{table*}
\centering
\caption{\textit{Expected RV dispersion from \citet{DehnenBinney1998} ($\sigma_{DB}$) compared to the measured RV dispersion ($\sigma_{178^{\circ}}$) for early A - early F stars (solar metallicity set). Also given are the typical HectoSpec measurement errors ($\epsilon_{RV}$) for the different subtype ranges, and the quadrature sum in combination with $\sigma_{DB}$. Finally, the excess $\sigma$ required to reconcile $\sigma_{DB}$ and $\sigma_{178^\circ}$ are provided. These excesses are compatible with the extra dispersion likely to be introduced by the presence of spectroscopic binaries in the sample (on the order of $10$\,km\,s$^{-1}$).}}
\label{tab:veldisp}
\begin{tabular}{cccccccc}
\textbf{Stellar Type} & \textbf{T$_{eff}$ range} & \textbf{No. stars} & \textbf{$\sigma_{DB}$} & \textbf{$\epsilon_{RV}$} & \textbf{$\sqrt{\sigma^2_{DB}+\epsilon^2_{RV}}$} & \textbf{$\sigma_{178^\circ}$} & \textbf{excess $\sigma$} \\ \hline
Early A               & {]}8500-10000{]}         & 67                 & 14    & 7.2                 & 15.7                              & 20.7                          & 13.5                                  \\
Late A                & {]}7500-8500{]}          & 151                & 19   & 5.5                  & 19.8                              & 21.7                          & 8.9                                   \\
Early F               & {]}6500-7500{]}          & 139                & 23    & 4.2                 & 23.4                              & 27.8                         & 15.0                                 \\
\hline
\end{tabular}
\end{table*}

\subsection{$\ell=118^{\circ}$}

\begin{figure*}
\centering
\includegraphics[width=\textwidth]{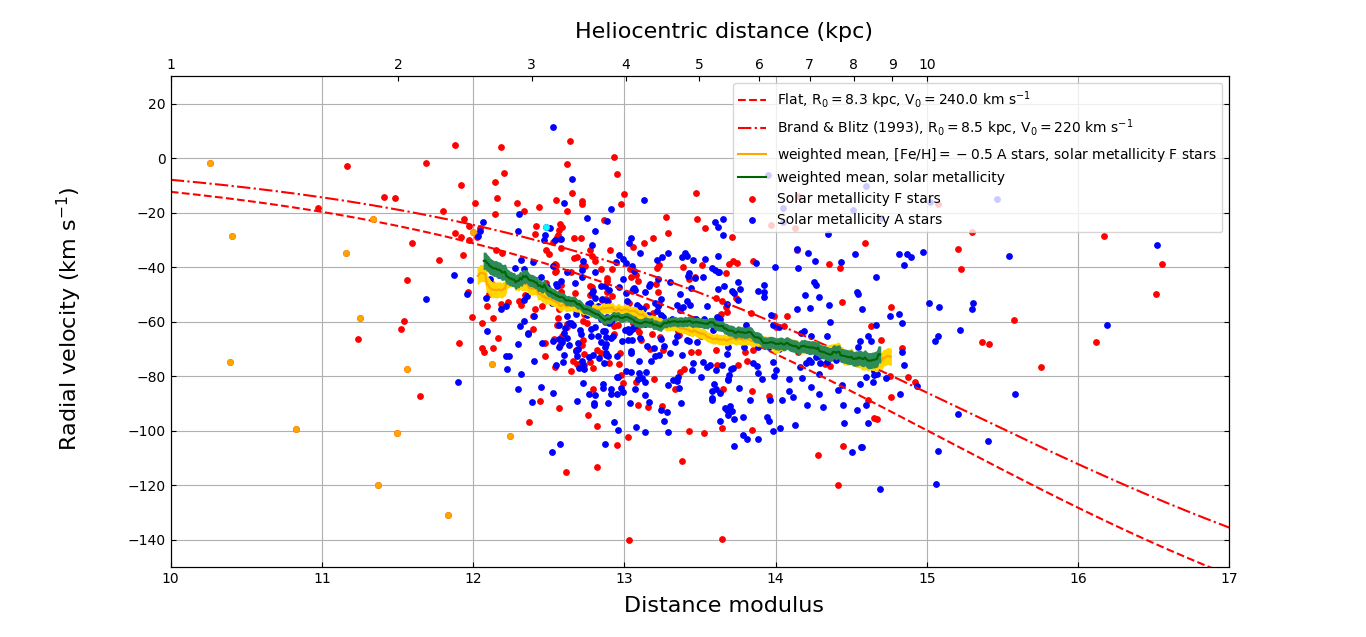}
\caption{\textit{The trend in RV with distance modulus at $\ell=118^{\circ}$. The red dashed/dotted lines show the expected trends for a flat rotation law and the \citet{BrandBlitz1993} slowly rising rotation law (see legend). The green line is the weighted mean of the RVs obtained using stellar parameters returned for solar metallicities, and the yellow line is the alternative result obtained on swapping in A star data computed for $\lbrack$Fe/H$ \rbrack = -0.5$. The shaded regions represent the standard error of the mean in RV. The individual data points in blue are obtained from A stars, while red is used for the F stars (computed for solar metallicity). The cyan point is the A star identified as lying far from the IPHAS mean extinction trend, and the orange points the F stars (see figure \ref{fig:extinction}). 
}}
\label{fig:rv_vs_dist_118}
\end{figure*}

The results for this sightline are shown in figure \ref{fig:rv_vs_dist_118}. Again, the green line is the mean trend for solar metallicity stars and the yellow for $\lbrack$Fe/H$\rbrack=-0.5$ A stars and solar metallicity F stars. The mean trends, calculated in the same way as for $\ell=178^\circ$ described in section \ref{sec:180}, deviate more strongly from expectations based on a flat or slowly rising rotation curve (shown in the figures by the red dashed/dotted lines). The measured trends are appreciably flatter.  The $\ell=118^{\circ}$ results are similar for both A star metallicities, indicating that metallicity is not having a strong effect on the interpretation of the results.  The mean RV trend (for either A-star metallicity) spans the heliocentric distance range from $2.5$ to $9$\,kpc (R$_G \simeq 9.7$ to $14.8$\,kpc), with the RV decreasing slowly from $\sim -40$ to $-70$\,km\,s$^{-1}$.


For context, we note that the Perseus Arm is likely to pass closest to this line of sight at a distance of 2.5 to 3\,kpc \citep{Reid2014}. This overlaps the bottom end of the distance range that we sample. In figure \ref{fig:rv_vs_dist_118} there is no clear sign of a distinct localised RV perturbation that might be attributed to the arm. Spiral arm perturbations are discussed further in section \ref{sec:spiral}.

The velocity dispersions around the RV mean trend for the three spectral type groups specified in table \ref{tab:veldisp} at the $\ell=118^{\circ}$ sightline are: 19.8\,km\,s$^{-1}$ for the early A stars, 21.1\,km\,s$^{-1}$ for the late A stars, and 24.3\,km\,s$^{-1}$ for the early F stars. These values are similar to those measured for the anticentre sightline (shown in table \ref{tab:veldisp}), which permits the same interpretation that binary orbital motions are a third factor contributing to the overall observed dispersion. 

In the next section, we consider whether the photometric consequences of binarity could distort the results, discuss the observed departure from a flat or slowly rising rotation law, and consider the possibilities for detecting spiral arm perturbations. 


\section{Discussion}
\label{sec:discussion}

\subsection{Potential bias from binaries and distance error}
The relatively flat distribution of RVs seen at $\ell=118^\circ$ in place of the expected falling trend may possibly be a consequence of unrecognised binaries placed at inappropriately short distances, or of significant smearing due to distance error.

If a target star is actually an unresolved binary system, the measured apparent magnitude will be brighter than if it were a single star. In effect, the absolute magnitude adopted in our analysis is then too low, and hence the distance will be underestimated. Similarly, if the component masses are unequal, the colour will be redder than if the system were a single star, resulting in overestimation of the extinction and further underestimation of distance. 
The additional velocity component from the motion around another star in a binary system does not produce a bias on the RV results since the space orientation of binary orbital axes across a large sample is random and must cancel out.
The photometric effect will be most pronounced in nearly equal mass binary systems, and has been quantified by \citet{HurleyTout1998}. Since the binary mass ratio distribution in A/F stars is not far from flat, we might expect of order $10-20\%$ of our sample to be misconstrued as closer to the Sun by up to $0.75$ in $\mu$. In principle this might erroneously flatten the overall RV distribution by bringing in a group of objects at more negative RV to mix with less negative values at smaller $\mu$.


In order to test the effect of binarity on the calculated weighted average RV trend, we have performed an outline simulation that focuses on this factor. Our method is as follows.  We select three sets of stars of different spectral type groups: early A, late A and early F. The size of the sets are the same as the HectoSpec $\ell=118^\circ$ groups (early A: 168 stars, late A: 281 stars, early F: 259 stars). 70\% of stars in each group are randomly selected as binaries -- these are assigned a primary mass, $m_1$, and the secondary mass, $m_2$, is assigned at random according to a mass ratio fraction, q, obeying the distribution $f(q)=q^{-0.5}$ \citep{Duchene2013}. The distance modulus assigned to each star is sampled from the HectoSpec $\mu$ distribution for its spectral type group. In order to compensate for the expected net decrease in $\mu$ associated with treating output binary stars as single, these reference distributions are first modified by a uniform retrospectively determined shift of +0.16. The assigned RVs follow a flat rotation curve with V$_0=240$\,km\,s$^{-1}$ at $\ell=118^{\circ}$, broadened by an amount consistent with the scatter defined by \citet{DehnenBinney1998} and measurement error. The final adjustment is to include a component of binary orbital motion. To achieve this, a period, P, inclination, i, and phase, $\phi$, must be randomly assigned for each binary star. The period is selected from the distribution of A star periods shown in figure 2 in \citet{Duchene2013}. The inclination is chosen from a uniform distribution in $\cos{i}$, and the phase from a uniform distribution between $0$ and $2\pi$. In the final reconstruction of the observed RV distribution, each simulated binary star has its distance modulus reduced to the equivalent single value according to the computed difference in intrinsic colour and absolute magnitude.

\begin{figure*}
\centering
\includegraphics[width=6in, keepaspectratio]{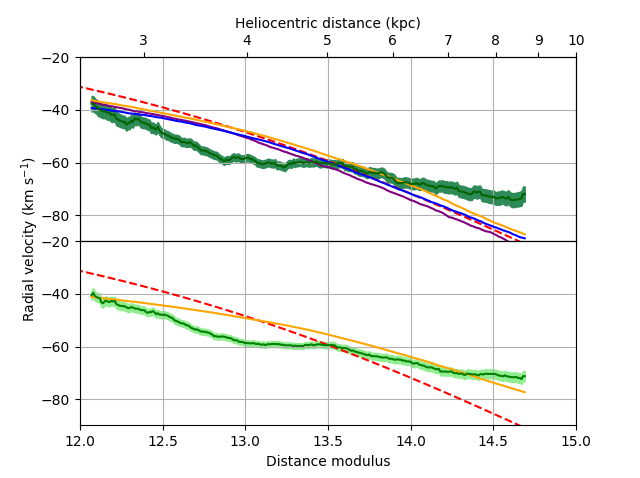}
\caption{\textit{The results of simulations to test the effects on the mean RV trend of undetected binaries and distance modulus errors.
In the top panel, the purple line is the mean RV trend for the simulation incorporating unresolved binaries. The orange line is the result from the simulation testing the effect of distance modulus error only, while the blue line shows the simulated effect of both unresolved binaries and distance modulus errors. The observed RV trend is overlaid in green. In the bottom panel, we test the effect of doubling our distance modulus errors (orange line).  The greater error induces more flattening.  The light green line in this case is the mean RV trend of the HectoSpec data obtained on adopting doubled distance modulus errors.}}
\label{fig:rv_dm_sim}
\end{figure*}

The simulation of 708 stars was performed 10000 times, and the mean RV trend was calculated each time. The mean of these 10000 trends is shown as the purple line in figure \ref{fig:rv_dm_sim}. The binary stars are pulled by varying amounts to shorter distances and bring with them their on-average more negative RV. A flattening of trend is seen, but our numerical experiment indicates it is modest. The deviation away from a flat rotation law we find at $\ell=118^\circ$ (figure \ref{fig:rv_vs_dist_118}) is much more pronounced, leading us to conclude that an appeal to stellar multiplicity to explain it falls well short, quantitatively. 

Another factor that will cause some flattening of the $\ell=118^\circ$ trend is distance error. The derived distance distibution is essentially the true distance distribution broadened by the uncertainty. Consequently the mean RV trend spans this slightly broader distribution, causing a flattening. To test the extent of this we performed a simulation similar to the one described above used to test the effect of unresolved binaries. However this time we disregard binary stars, and after assigning a suitably scattered RV with measurement error to each notional star, we shifted the corresponding distance modulus by an amount within the typical level of error from the HectoSpec data. The result is shown as the orange line in figure \ref{fig:rv_dm_sim}. The resultant mean trend is somewhat flattened, but not by an amount that makes it parallel to the result from observation.  To complete the picture, we performed this simulation again but now including unresolved binaries as described above. The blue line in figure \ref{fig:rv_dm_sim} shows the result. The amount of flattening is similar to that due to the distance errors alone. At shorter distances it is slightly more pronounced, and at further distances it is less. This is expected since the binaries are pulled to shorter distances, bringing with them their on-average more negative RV.

In order to achieve a flattening of the simulated curve by an amount that begins to mimic the trend deduced from the A/F star data, we find we need to double the distance errors relative to those propagated from the data as described in section~\ref{sec:mag_ex_dis}.  The second panel of figure~\ref{fig:rv_dm_sim} illustrates this.  Whilst it is certainly a possibility the estimation of distance errors in our sample is optimistic at $\sim$ 15\% ($\Delta \mu = 0.3$), growing them all by as much as a factor of two is rather less credible.

\subsection{Comparisons of results with other tracers and earlier work}
\label{sec:comparison}
Using H\,\textsc{ii} region data, \citet{BrandBlitz1993} presented sparse-sampled measurements that mainly captured heliocentric distances out to $\sim$4 kpc at the Galactic longitudes of interest here (see their figure 1).  The overlap with our results thus runs roughly from 2 to 4 kpc.   Near the anticentre \citet{BrandBlitz1993} generally favoured slightly negative radial velocities, ranging from $-18$ to $+8$ km s$^{-1}$ (9 datapoints, from their table 1), to be compared with a small positive bias here (figure~\ref{fig:rv_vs_dm}).   In the longitude range $110^{\circ} < \ell < 130^{\circ}$, the relevant measurements are spread between $-30$ and $-56$ km s$^{-1}$ (13 datapoints).  This is entirely compatible with our results.



A denser comparison between our results and other studies can be made using RV data of H\,\textsc{i} and CO clouds. Since both sightlines miss the latitude of peak gaseous emission for their longitudes, the total amount of both H\,\textsc{i} and CO is not particularly large. Nevertheless, the measured gaseous RV distributions are broadly consistent with our findings from HectoSpec, and the detail is informative.  H\,\textsc{i} 21\,cm data from EBHIS \citep[The Effelsberg-Bonn H\,\textsc{i} Survey,][]{Winkel2016} in the $\ell=178^\circ$ sightline scatter around RV $\simeq$ 0 at much reduced dispersion compared to the A/F stars -- as expected. We find that the mean RV measures from our optical spectroscopy are shifted relative to the H\,\textsc{i} data by $\sim +8$\,km\,s$^{-1}$, and regard this mainly as evidence that the H\,\textsc{i} column samples a greater column through the outer disc than our stellar sample.  

\begin{figure}
\centering
\includegraphics[width=\columnwidth]{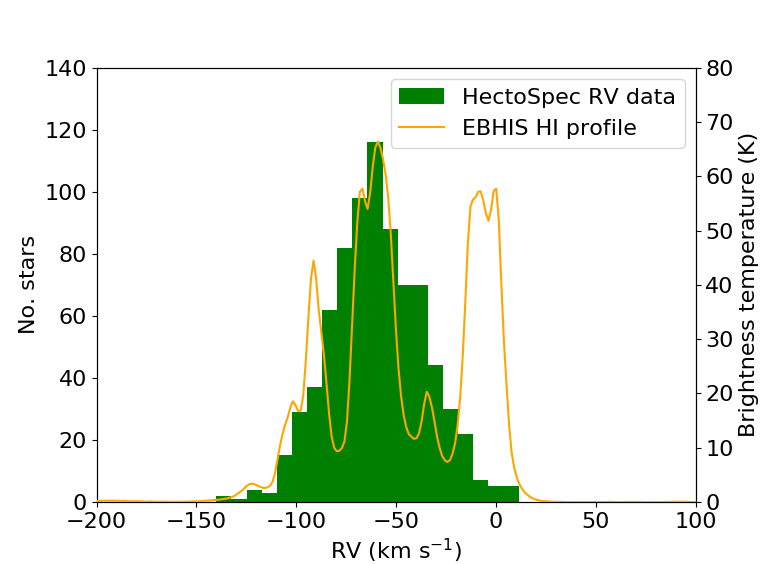}
\caption{\textit{The H\,\textsc{i} profile at $\ell=118^\circ$ (orange line), overlayed on the HectoSpec RV data (green histogram).}}
\label{fig:HI}
\end{figure}

The comparison with EBHIS data at $\ell = 118^{\circ}$ shown in figure \ref{fig:HI} shows the A/F star data line up with the main $\sim -60$\,km\,s$^{-1}$ H\,\textsc{i} emission peak, while the peak at $\sim 0$\,km\,s$^{-1}$ (from the Local Arm, seen in H\,\textsc{i}) is clearly absent.  This is to be expected given that none of the A/F stars selected and measured will be in or near the Local Arm.  The H\,\textsc{i} data also present a peak at $\sim -90$\,km\,s$^{-1}$ that is largely absent from the stellar data. This is unsurprising since our central result from the $\ell = 118^{\circ}$ sightline is the RV flattening that implies a relative absence of stars in this more negative velocity range (at distances where a flat or gently rising rotation law would predict they exist).  A reasonable inference from this is that the H\,\textsc{i} gas at the most negative radial velocities lies mainly outside the range sampled by the A/F stars.  We note that the CO data from the COMPLETE \citep[Coordinated Molecular Probe Line Extinction and Thermal Emission][]{Ridge2006} survey exhibit the same RV peaks as the H\,\textsc{i} data. This difference either indicates that the H\,\textsc{i} and CO gas lies beyond $d \sim 9$\,kpc, or that the distance range occupied by our sample of stars does not extend to the distances where existing rotation laws would predict RV$\sim-90$\,km\,s$^{-1}$. From figure \ref{fig:rv_vs_dist_118}, it can be seen that these laws associate a distance of $\sim 8$\,kpc with this RV. We argue below the stars in our sample placed at $\sim 8$\,kpc are not there simple because of distance error.

\begin{figure*}
\centering
\includegraphics[width=\textwidth]{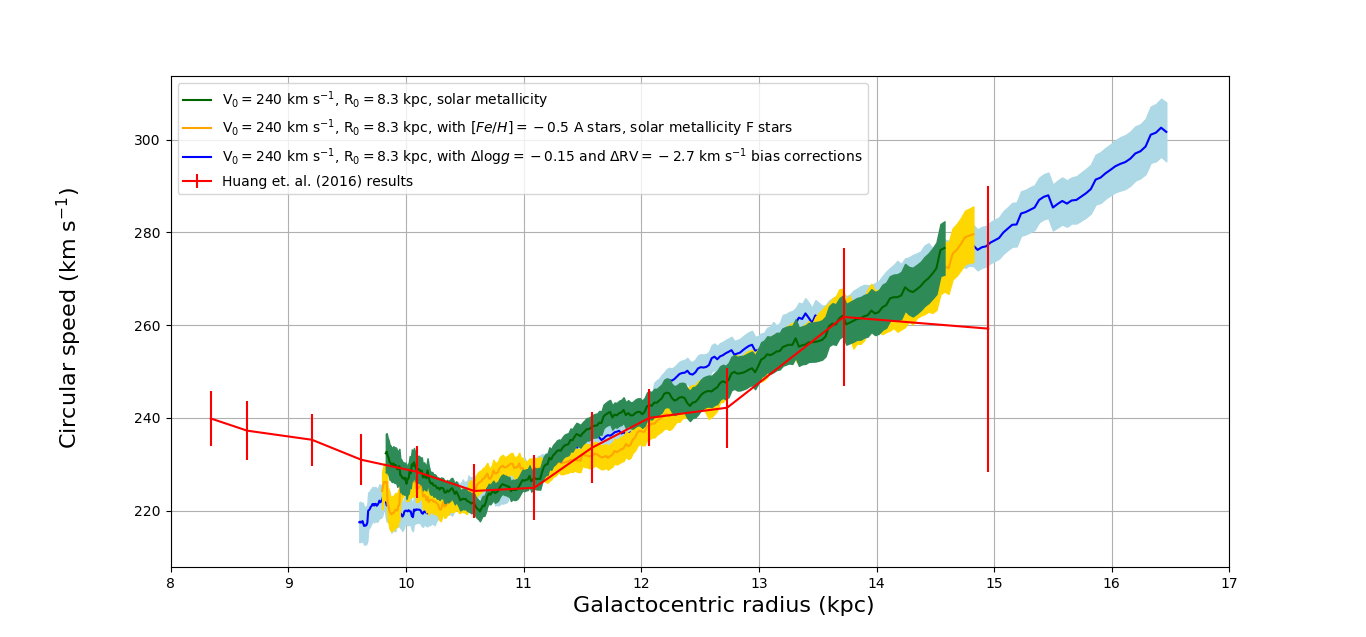}
\caption{\textit{ Galactic disc circular speed results from \citet{Huang2016} are reproduced in red. The circular speeds derived from the mean RV trend for results at $\ell=118^\circ$, adopting solar metallicity, are in green.  Results obtained with $\Delta \log{g}=-0.15$ and $\Delta$RV$=-2.7$\,km\,s$^{-1}$ potential bias corrections are shown in blue.  Orange is used for the results obtained when the A star parameters for $\lbrack$Fe/H$ \rbrack =-0.5$ are used in place of solar metallicity parameters. The shaded region around each HectoSpec line represents its error -- propagated from the error on the mean trend in figure \ref{fig:rv_vs_dist_118}.}}
\label{fig:rotcurve}
\end{figure*}

\citet{Huang2016} have used red clump giants drawn from a wide range of Galactic longitudes sampling the outer disc to find a broadly flat longitude-averaged rotation law within R$_G < 25$\,kpc, with typical circular speed V$_0=240\pm6$\,km\,s$^{-1}$. But over our sampled region, between R$_G \simeq 10-15$\,kpc, their inferred rotation law is quite sharply rising out of a dip at R$_G \sim 11$\,kpc. Figure \ref{fig:rotcurve} compares the \citet{Huang2016} results (red line) with the rotation curve derived from the mean RV trend we obtain at $\ell=118^\circ$ adopting solar metallicity (green line). In constructing this figure, we choose the same LSR parameters as favoured by \citet{Huang2016}, R$_0=8.3$\,kpc and V$_0=240$\,km\,s$^{-1}$. The agreement is very good. 

We commented before in sections \ref{sec:logg} and \ref{sec:isis} that there may be a positive bias in the derived stellar surface gravities. The amount of bias may very well be in the region of $\Delta \log{g}=+0.15$, as suggested by figure \ref{fig:logg}. The remeasurements of just 7 objects find in favour of a larger bias (see section \ref{sec:isis}), but we do not otherwise see evidence that the distance range of our stellar sample, strongly influenced by $\log{g}$, is significantly underestimated (cf. figure \ref{fig:extinction} and the linked discussion). We have also considered whether there is a bias in the measured RV. The best evidence we have of this comes from the 7 remeasurements at red and blue wavelengths: the weighted mean offset obtained is $-2.7$\,km\,s$^{-1}$. If all stellar surface gravities are corrected down by 0.15 dex, and this potential RV bias is also taken out uniformly, the trend in circular speed aquires the form of the blue line in figure \ref{fig:rotcurve}. The main impact of these changes is to stretch the results out to an increased Galactocentric radius, in response to the $\log{g}$ shift. The outcome remains consistent with the \citet{Huang2016} results. Another factor playing a role is the presumed metallicity for the A stars: the orange line in the figure shows the rotation curve when setting $\lbrack$Fe/H$ \rbrack = -0.5$ for the A stars.  It can be seen that this change also has little impact.

So we have that both the clump giants and, now, our A/F star sample favour a rotation law that rises out to $R_G \sim 14$ kpc, after a minimum near $\sim 11$ kpc.  But we have also demonstrated how distance error can modify the observed RV-distance trend.  And indeed \citet{BinneyDehnen1997} presented a thought experiment that drew attention to how a linear increase in Galactic rotation into the outer disc would arise this way. The particular example they presented was of the inferred law from tracers confined within a ring -- mimicking gas tracers associated with spiral arms -- at 1.6 times the Sun's Galactocentric radius.  These were subject to distance uncertainties similar to the larger errors considered in the lower panel of figure \ref{fig:rv_dm_sim}.  This extreme is avoided here.  The facts of the spread in stellar parameters ($\Delta M_i \sim 2$) and apparent magnitudes ($\Delta m_i \sim 1.5$, see figure~\ref{fig:count_vs_i}) combine with gently rising extinction (see figure~\ref{fig:extinction}) to yield an underlying stellar distribution that should span at least 5 kpc -- nor is there an expectation these stars would be confined to e.g. just the Perseus Arm.  This leaves us cautiously supporting the case for an increase in circular speed outside the Solar Circle and looking forward to the more extensive studies needed to clarify the situation.      

\subsection{Spiral arm perturbations}
\label{sec:spiral}

Spiral arms in galactic discs are widely viewed as linked with non-axisymmetric kinematic perturbations. To assess whether our data can expose such an effect, we examine the $\ell=178^\circ$ sight line, since it is close to the radial direction that minimises shear due to Galactic rotation and more easily reveals low-amplitude perturbations that may be associated with spiral arm structure. \citet{Monguio2015} used B4-A1 stars to find a stellar overdensity due to the Perseus spiral arm at a heliocentric distance of $1.6\pm0.2$\,kpc in the anticentre direction, and \citet{Reid2014} used parallaxes of 24 star forming regions to find the arm to be located at $2$\,kpc. The Perseus Arm is therefore located just short of the sampled region in this work. However \citet{Reid2014} also found evidence that an Outer Arm is located at a heliocentric distance of $6$\,kpc in the anticentre direction. This arm lies near the far end of our sampled region. 

The scale of radial velocity perturbation depends on the model adopted for the origin of the perturbation. On the one hand, \citet{Monari2016} simulated the effect of a spiral potential on an axisymmetric equilibrium distribution function (emulating the Milky Way thin stellar disc) and found radial velocity perturbations of order $-5$\,km\,s$^{-1}$ within the arms and $+5$\,km\,s$^{-1}$ in between them. On the other, \citet{Grand2016} favour the transient winding arm view and find the perturbation of young stars ($<3$\,Gyr) to be considerably stronger at up to $20$\,km\,s$^{-1}$. 

We do not see any clear signs of perturbations, negative or positive, alligned with the Outer Arm in the HectoSpec results. However since the HectoSpec RV error is comparable with predicted  perturbations at the lower end of the expected range, and the stellar sample is subject to distance error, it is not obvious that we should expect to observe their signal in our data. In order to test this, we conducted another simulation. A sample of stars the same size as the HectoSpec $\ell=178^\circ$ sample and with the same distance distribution were assigned RVs according to a sinusoidal waveform -- three separate tests were conducted with amplitudes of 5, 10 and 20\,km\,s$^{-1}$. Two spiral arms were put at 2\,kpc and 6\,kpc. Velocity scatter, RV error and distance error were then applied to the distribution, and a mean trend was computed as in section \ref{sec:180}. 

The results obtained for the small perturbations (5\,km\,s$^{-1}$) exhibited no clear signs of the input sinusoid's phasing or wavelength, giving a mean trend compatible with zero for most of the sampled range. In contrast, the results for the larger amplitude perturbations did show signs of the input phasing, wavelength and amplitude. 
We conclude that spiral arm perturbations of small amplitude would be unlikely to appear with clear statistical significance in our results, but those of larger amplitude would.


Returning to the HectoSpec $\ell=178^\circ$ results themselves (figure \ref{fig:rv_vs_dm}), we do observe a wave-like structure in the mean RV trend with amplitude $\sim5-10$\,km\,s$^{-1}$, but not at an implied phase or long-enough wavelength that would make sense in comparison with the expected locations of the Perseus Arm and Outer arms (at $\sim2$\,kpc and $\sim6$\,kpc in this sightline, respectively).  If the RV wobble is real, rather than a sample size effect, it is most likely a local effect unconnected with the larger scale structure of the Galactic disc. But we must discard the possibility of large amplitude (10-20\,km\,s$^{-1}$) perturbations in this direction.  The results for this sightline are in keeping with the findings of \citet{Fernandez2001}, who used both OB stars and Cepheids to limit perturbations to under 3\,km\,s$^{-1}$.

Finally we comment on the form of the $\ell=118^\circ$ results (figure~\ref{fig:rv_vs_dist_118}). Specifically, can a spiral arm perturbation explain the observed deviation from the trend predicted by a flat rotation law? To make this comparison, the same three simulations as described above for the $\ell=178^\circ$ sightline were conducted for $\ell=118^\circ$. In this case the spiral arms are located at $3$ and $6.5$\,kpc \citep{Reid2014} and account is taken of the expected combination of Galactic azimuthal and radial perturbations along the line of sight. In this case the observed RV trend, when compared with the simulation alternatives, shows a deviation from the flat rotation law of a scale similar to that of the largest ($20$\,km\,s$^{-1}$) perturbation investigated. This stands in clear contrast to the low-amplitude perturbation compatible with the $\ell=178^\circ$ sightline. Hence it appears there is no simple coherent way of interpreting our radial velocity trends in terms of one class of spiral arm perturbation model.

\section{Conclusions}
\label{sec:conclusion}

For the first time, we have demonstrated that A/F star spectroscopy, even when restricted to a small region around the Calcium triplet lines, can provide useful insights into the kinematics of the Milky Way disc. With the use of the method employed in this paper, stellar parameters and radial velocities can be measured for the very large samples of these stars, with $i < 18$ (Vega), that will be accessible to future massive-multiplex instruments such as the WEAVE and 4MOST spectrographs being constructed for the William Herschel and VISTA telescopes respectively. 

This first result, deploying a dense sample of $\sim$800 A/F stars within just one pencil beam, 1 degree in diameter, at $\ell = 118^{\circ}$, fits in well with the trend of a rising mean circular speed within the Galactic disc reported recently by \citet{Huang2016}, over the Galactocentric radius range $11 <$  R$_G$ (kpc) $< 15$. Interpreting these RV measurements in terms of spiral arm perturbation is more difficult as then it becomes hard to reconcile the $\ell=118^{\circ}$ and $\ell = 178^\circ$ results. The \citet{Huang2016} sample of clump giants is much larger, comprising $\sim$16000 stars, drawn from a very broad fan of outer-disc longitudes ($100^{\circ} < \ell < 230^{\circ}$, roughly) -- hence their results describe a longitude average, in contrast to ours.  Further dense-sampling studies like ours will be needed to discover whether the rotation law is closely axisymmetric as has usually been assumed, hitherto.  This first comparison passes the test. However, both stellar results are at variance with the maser-based study of \citet{Reid2014} which obtained evidence of a very nearly flat rotation law to $R_G = 16$ kpc, as well as with the long-established \citet{BrandBlitz1993} law. The origin of this difference may rest in the very limited maser data at all longitudes beyond R$_G=12$\,kpc \citep[see figure 1 of][]{Reid2014}. On the other hand, caution still needs to be exercised that some of the inferred steepening may prove to be a consequence of uncertainties in stellar distances derived from spectroscopic parallax. If the steepening is correct, it would lend support to previous evidence of a ring of dark matter in the outer disc at R$_G$ between 13 and 18\,kpc \citep[e.g see][]{Kalberla2007}. 

Our goal is to open up the use of A/F stars in studies of Galactic disc structure at red wavelengths that mitigate the effects of extinction. It is more common to derive stellar parameters for these earlier type stars in the blue optical and so it was not evident at the outset whether fitting to the red range only would suffer from biases or degeneracies in parameter determination.  For the present purpose, it has become clear that degeneracies are less troublesome for A stars than F stars, although it would be desirable in future to better disentangle effective temperature and metallicity, even for A stars -- better spectral resolution and wider wavelength coverage can both help here.  We plan to examine these options.  The basic method of target selection, using IPHAS colours, is certainly very efficient \citep[see also][]{Hales2009}.  We have assessed possible bias in our numerical results: whilst there appears to be some, our tests examining their impact have so far shown that the impact is modest. Furthermore, we find that the random error on the mean trend in circular speed with increasing Galactocentric radius, achieved with our dense localised sampling, is generally better than $\pm 5$\,km\,s$^{-1}$, and represents an advance on previous measurements. 

Since the single biggest source of uncertainty in a study like this originates in the stellar distances, we look forward to the new opportunity that starts to take shape with the stellar astrometry in the Gaia DR2 release in April 2018.  A representative $i$ magnitude among our selected A/F stars is $\sim 17$ (or Gaia $G$ approaching 18): at this brightness, the anticipated DR2 (end-of-mission) parallax and proper motion errors are likely to be 150 (100) $\mu$as, and 80 (50) $\mu$as yr$^{-1}$, respectively \citep[see ESA website and][]{Katz2017}.  Whilst distances can not be nailed down for individual stars, over the range of interest here from $\sim$2 to almost 10 kpc (parallaxes of 500 down to 100 $\mu$as), significantly improved constraints will become available.  There will also be the opening to bring into account full space motions, on folding in Gaia proper motion data.  As our next step to prepare for this, we are building customised forward simulations based on Galactic models aimed at predicting full space motion distributions.  As part of this we will examine in further detail the roles played by our methods of target selection and analysis.  We have already checked one of the more obviously significant issues that concerns us - no weeding of binary stars - and have found that to be a small effect. 

Another experiment for the future will be to examine the differences in performance of A-star and clump-giant samples, on the same sightline.  Based on what we have learned here, there is reason to expect A-star samples to be especially valuable in characterising the outer Galactic disc, where these younger less-scattered stars are relatively more common.

\section*{Acknowledgments}
AH and HF acknowledge support from the UK's Science and Technology Facilities Council (STFC), grant nos. ST/K502029/1 and ST/I505699/1.  JED and MM acknowledge funding via STFC grants ST/J001333/1 and ST/M001008/1. JJD was funded by NASA contract NAS8-03060 to the {\it Chandra X-ray Center}. ~SES acknowledges support via STFC grant no. ST/M00127X/1. NJW acknowledges an STFC Ernest Rutherford Fellowship, grant no.~ST/M005569/1.

Observations reported here were obtained at the MMT Observatory, a joint facility of the Smithsonian Institution and the University of Arizona. This study has also benefitted from the service programme (ref. SW2016b05) on the William Herschel Telescope (WHT): the WHT and its service programme are operated on the island of La Palma by the Isaac Newton Group of Telescopes in the Spanish Observatorio del Roque de los Muchachos of the Instituto de Astrofísica de Canarias.    

We thank the referee of this paper for useful comments that have helped us improve the paper's content.

\bibliographystyle{mn2e}
\bibliography{write_up_bib5}

\appendix

\label{lastpage}

\end{document}